\documentclass[prl,aps,superscriptaddress,preprint,floatfix,
nofootinbib]{revtex4-1}
\usepackage[utf8x]{inputenc}

\usepackage{amsmath,amsfonts,amssymb}
\usepackage{graphicx}
\usepackage{dcolumn}
\usepackage{bm}
\usepackage{subfigure}

\usepackage{color}
\usepackage{dsfont}

\begin{document}

\title{Unconventional Chiral Fermions and Large Topological Fermi Arcs in RhSi}
\author{Guoqing Chang}
\thanks{These authors contributed equally to this work.}
\affiliation{Centre for Advanced 2D Materials and Graphene Research Centre National University of Singapore, 6 Science Drive 2, Singapore 117546}\affiliation{Department of Physics, National University of Singapore, 2 Science Drive 3, Singapore 117542}

\author{Su-Yang Xu}
\thanks{These authors contributed equally to this work.}
\affiliation {Laboratory for Topological Quantum Matter and Spectroscopy (B7), Department of Physics, Princeton University, Princeton, New Jersey 08544, USA}
\thanks{Corresponding authors (emails): suyangxu@princeton.edu, nilnish@gmail.com, mzhasan@princeton.edu }

\author{Benjamin J. Wieder}
\thanks{These authors contributed equally to this work.}
\affiliation{Nordita, Center for Quantum Materials, KTH Royal Institute of Technology and Stockholm University, Roslagstullsbacken 23, SE-106 91 Stockholm, Sweden}
\author{Daniel S. Sanchez}
\thanks{These authors contributed equally to this work.}
\affiliation {Laboratory for Topological Quantum Matter and Spectroscopy (B7), Department of Physics, Princeton University, Princeton, New Jersey 08544, USA}
\author{Shin-Ming Huang}
\affiliation{Department of Physics, National Sun Yat-sen University, Kaohsiung 804, Taiwan}
\author{Ilya Belopolski}\affiliation {Laboratory for Topological Quantum Matter and Spectroscopy (B7), Department of Physics, Princeton University, Princeton, New Jersey 08544, USA}
\author{Tay-Rong Chang}
\affiliation{Department of Physics, National Cheng Kung University, Tainan, 701, Taiwan}
\author{Songtian Zhang}\affiliation {Laboratory for Topological Quantum Matter and Spectroscopy (B7), Department of Physics, Princeton University, Princeton, New Jersey 08544, USA}
\author{Arun Bansil}
\affiliation{Department of Physics, Northeastern University, Boston, Massachusetts 02115, USA}
\author{Hsin Lin}
\thanks{Corresponding authors (emails): suyangxu@princeton.edu, nilnish@gmail.com, mzhasan@princeton.edu }
\affiliation{Centre for Advanced 2D Materials and Graphene Research Centre National University of Singapore, 6 Science Drive 2, Singapore 117546}
\affiliation{Department of Physics, National University of Singapore, 2 Science Drive 3, Singapore 117542}
\author{M. Zahid Hasan}
\thanks{Corresponding authors (emails): suyangxu@princeton.edu, nilnish@gmail.com, mzhasan@princeton.edu }
\affiliation {Laboratory for Topological Quantum Matter and Spectroscopy (B7), Department of Physics, Princeton University, Princeton, New Jersey 08544, USA}

\begin{abstract}
The theoretical proposal of chiral fermions in topological semimetals has led to a significant effort towards their experimental realization.  In particular, the Fermi surfaces of chiral semimetals carry quantized Chern numbers, making them an attractive platform for the observation of exotic transport and optical phenomena.  While the simplest example of a chiral fermion in condensed matter is a conventional $|C|=1$ Weyl fermion, recent theoretical works have proposed a number of unconventional chiral fermions beyond the Standard Model which are protected by unique combinations of topology and crystalline symmetries.  However, materials candidates for experimentally probing the transport and response signatures of these unconventional fermions have thus far remained elusive.  In this paper, we propose the RhSi family in space group (SG) $\#$198 as the ideal platform for the experimental examination of unconventional chiral fermions.  We find that RhSi is a filling-enforced semimetal that features near its Fermi surface a chiral double six-fold-degenerate spin-1 Weyl node at $R$ and a previously uncharacterized four-fold-degenerate chiral fermion at $\Gamma$. Each unconventional fermion displays Chern number ±4 at the Fermi level. We also show that RhSi displays the largest possible momentum separation of compensative chiral fermions, the largest proposed topologically nontrivial energy window, and the longest possible Fermi arcs on its surface. We conclude by proposing signatures of an exotic bulk photogalvanic response in RhSi.
\end{abstract}

\maketitle

The allowed band crossings in condensed matter have, until recently, been considered closely linked to elementary particles in high-energy physics~\cite{Weyl, Wilczek,Volovik2003, NewFermion}.  In 3D systems without spatial inversion ($\mathcal{I}$) \emph{or} time-reversal ($\mathcal{T}$) symmetry, two-fold-degenerate band crossings are permitted, resulting in condensed matter realizations of Weyl fermions with quantized Chern numbers~\cite{Weyl, Wilczek,Volovik2003, NewFermion, Kane05p226801,Kane05p146802,Hasan2010,Qi2011,BernevigHgTe,RevBansil, Hasan_TaAs, TaAs_Ding, MIT_Weyl, Wan2011,Burkov2011,Murakami2007,Huang2015, Weng2015, type2Weyl1,type2Weyl2, LaAlGe, KramersWeyl}.  Since the experimental realization of the Weyl semimetal state in TaAs~\cite{Hasan_TaAs, TaAs_Ding}, recent theoretical efforts have become focused on finding unconventional condensed matter quasiparticle excitations beyond the Dirac and Weyl paradigm described by the Standard Model~\cite{DoubleDirac,NewFermion,trip1,trip2,Nexus,Phonon}.  These efforts have rapidly expanded the set of known nodal features, which now additionally include symmorphic three-fold nexus fermions~\cite{trip1,trip2,Nexus}, eight-fold-degenerate double Dirac fermions~\cite{DoubleDirac}, and, as detailed by Bradlyn, Cano, Wang, \emph{et al.} (BCW) in Ref.~\onlinecite{NewFermion}, three-fold-degenerate single and six-fold-degenerate double spin-1 Weyl points.  Unconventional chiral fermions, in particular, hold great promise for experimental applications, as they broaden beyond conventional Weyl semimetals the search for materials candidates for the observation of topological surface states, bulk chiral transport, and exotic circular photogalvanic effects~\cite{MMonFS, chiral_mag, photocurrentweyl, photoTaAs, photogalvanic, arc1,arc2,Riemann}.

Although the fundamental theory for these unconventional fermions has been established, one outstanding issue has been the relative lack of ideal material candidates for their experimental examination.  In the band structures of previously proposed materials, the unconventional fermions have typically sat away from the Fermi energy, or have in the cases of unconventional chiral fermions coexisted with additional, trivial bands.  In these systems, while the unconventional fermions may be experimentally observable by photoemission, their topological properties are still prohibitively difficult to detect and utilize for transport and optical response.  For example in MoP, a three-fold nexus fermion is observed 1 eV below the Fermi level, but the Fermi surface itself is unrelated to the three-fold fermion and carries no net Chern number~\cite{Nexus_exp}.

\begin{figure}[t]
\includegraphics[width=140mm]{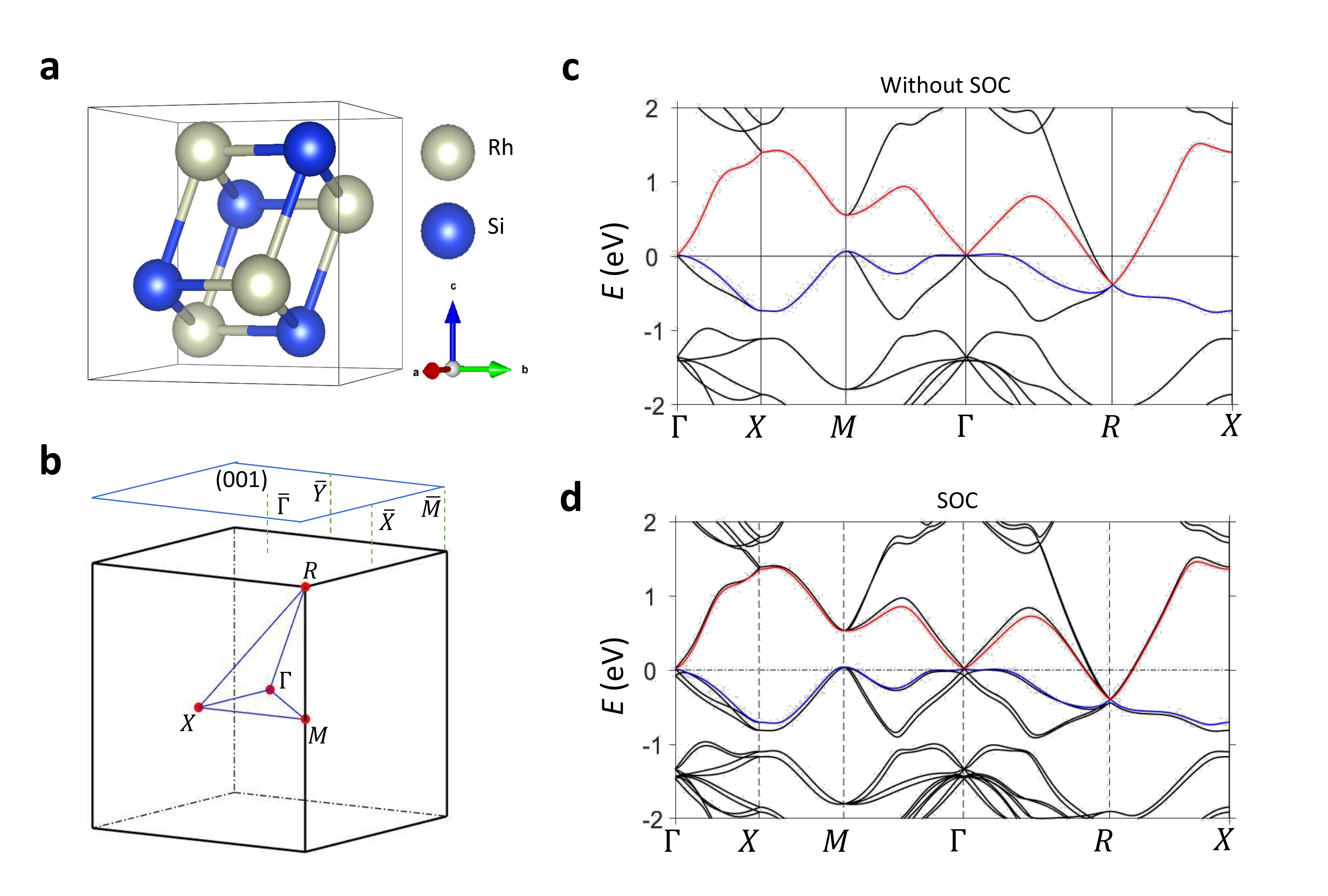}
\caption{\textbf{Lattice and electronic structure of RhSi in SG 198}. (a) Crystal structure of RhSi. Each unit cell contains 4 Rh and 4 Si atoms lying at Wyckoff positions with the minimum multiplicity of SG 198. (b) The cubic bulk Brillouin zone (BZ) of RhSi. (c) Band structure of RhSi in the absence of spin-orbit coupling (SOC). The highest valance and lowest conduction bands are colored in blue and red, respectively. (d) Band structure in the presence of SOC.  A chiral double spin-1 Weyl point sits $\sim0.4$eV below the Fermi energy at $R$ and a previously uncharacterized four-fold-degenerate chiral fermion lies at the Fermi energy at $\Gamma$.}
\label{Fig1}
\end{figure}

In this paper, we identify the RhSi materials family of structurally chiral cubic crystals in space group (SG) 198 $P2_{1}3$~\cite{RhSi} (Fig.~\ref{Fig1}(a)) as the first ideal materials candidates for the experimental study of the novel transport and response effects of unconventional chiral fermions. Using first-principles calculations detailed in Section A of the Supplemental Material (SM A), we find that the Fermi surface of RhSi consists of only two well-isolated pieces which carry equal and opposite quantized Chern number.  The bulk bands near the Fermi energy feature a chiral six-fold-degenerate double spin-1 Weyl at the Brillouin zone (BZ) corner $R$ and a previously uncharacterized four-fold-degenerate chiral fermion at the zone center $\Gamma$ (Fig.~\ref{Fig1}(c,d), Fig.~\ref{Fig2}(a,b)).  RhSi therefore displays the largest possible separation of chiral fermions allowed in crystals.  With an otherwise large bandgap, RhSi also therefore features the largest topologically nontrivial energy window proposed thus far (Fig.~\ref{Fig1}(d) and SM E).  Furthermore, as these two chiral fermions lie at time-reversal-invariant momenta (TRIMs), they are unrelated by symmetry and free to exhibit an energy offset; here, the four-fold fermion at $\Gamma$ lies roughly 400 meV above the six-fold fermion at $R$.  This offset allows for the possibility of unique optical transport, such as the quantized circular photogalvanic effect~\cite{photogalvanic}.  Among all known chiral semimetals, both conventional Weyl and unconventional higher-fold fermion, RhSi therefore stands as possibly the most electronically ideal material yet proposed.

To understand the unusual high-fold-degenerate nodes displayed in the minimal band connectivity of SG 198, we construct an eight-band tight-binding (TB) model (SM C).  SG 198 is characterized by three nonintersecting two-fold screw rotations $s_{2x,y,z}$, related by diagonal cubic three-fold rotation $C_{3,111}$~\cite{BigBook}:
\begin{eqnarray}
s_{2x}=\left\{C_{2x}\bigg|\frac{1}{2}\frac{1}{2}0\right\},\ &s_{2y}&=\left\{C_{2y}\bigg|0\frac{1}{2}\frac{1}{2}\right\}\nonumber \\
s_{2z}=\left\{C_{2z}\bigg|\frac{1}{2}0\frac{1}{2}\right\},\ &C_{3,111}&=\left\{C_{3,111}\bigg|000\right\}.
\label{gen198}
\end{eqnarray}
Without the three-fold rotation, this combination of screws and $\mathcal{T}$-symmetry characterizes orthorhombic SG 19, and has been shown to force groups of eight or more bands to tangle together~\cite{fillingconstraint,Layer_group,QuantumChemistry,Adrian19,BalatskySG19}.  The additional cubic three-fold rotation $C_{3,111}$ in SG 198 serves to increase the band degeneracy at TRIMs while still preserving this eight-band connectivity.  At an electron filling of $\nu\in8\mathbb{Z} + 4$, RhSi is gapless due to the combination of time-reversal and nonsymmorphic symmetries, and is therefore a ``filling-enforced'' semimetal (SM B)~\cite{fillingconstraint,Layer_group,magDirac2}.  We find that our minimal TB model of SG 198 captures all of the degeneracy structure and topological character of RhSi.  We describe our results for the full BZ in detail in SM C, and here focus on the chiral node structure at $\Gamma$ and $R$.

We begin by examining the band splitting and \textit{previously uncharacterized} four-fold-degenerate unconventional chiral fermion at $\Gamma$. In the absence of SOC, our eight-band model permits only a single mass term at $\Gamma$ which splits bands into a $3\times2$-fold-degenerate fermion and a doubly degenerate quadratic band, which in RhSi lies more than 2 eV above the Fermi energy (Fig.~\ref{Fig1}(c)).  Upon the introduction of SOC, this quadratic crossing opens into a Kramers Weyl~\cite{KramersWeyl}, and the $3\times2$-fold-degenerate node splits into a four-fold-degenerate unconventional fermion and a second Kramers Weyl (Fig.~\ref{Fig2}(a,d)).  This four-fold-degenerate fermion is distinct from the spin-3/2 chiral fermion introduced in Ref.~\onlinecite{NewFermion}: whereas that fermion is described by a corepresentation equivalent to the four-dimensional irreducible representation $\bar{F}$ of chiral point group $432$ ($O$), the four-fold-degenerate fermion in RhSi is described by the $\mathcal{T}$-symmetric corepresentation formed by pairing the two-dimensional irreducible representations $^{1}\bar{F}$ and $^{2}\bar{F}$ of chiral point group $23$ ($T$)~\cite{BigBook}.  In the language of atomic orbitals, this four-fold degeneracy can be understood by modeling the six degenerate states without SOC by three $p$ orbitals and an electron spin in the $111$ direction.  Calling $z'$ the $111$ direction and $x',y'$ as orthonormal axes spanning the plane normal to $z'$, we group the $p$ orbitals into a $p_{z'},\ m_{l}=0$ orbital and $p_{x'}\pm i p_{y'},\ m_{l}=\pm 1$ orbitals.  When coupled to the spin-$1/2$ electron, the six total states split into four $J=1/2$ and two $J=3/2$ states.  Time-reversal pairs states with the same $J$ and opposite $m_{j}$, and $s_{2x}$ flips $m_{s}$ without affecting $m_{l}$, such that under the SG 198 generators two $J=3/2$ states pair with two $J=1/2$ states and the remaining two $J=1/2$ states split off and form the second Kramers Weyl (SM C.2).  By numerically calculating the eigenvalues of $C_{3,111}$ and considering the symmetry-allowed term $\vec{k}\cdot\vec{J}$, each band near $\Gamma$ can be assigned $J$ and $m_{j}$ eigenvalues, a structure we confirm explicitly with a symmetry-generated four-band $k\cdot p$ model in SM C.2.  As the irreducible representations at $\Gamma$ are reflective of the position-space atomic orbitals, this analogy should also provide physical insight into the bonding character of RhSi~\cite{QuantumChemistry}.  By the integrating the Berry curvature between bands with $J=1/2,\ m_{j}=\pm 1/2$ over a k-space sphere in the vicinity of $\Gamma$~\cite{Volovik2003}, we find that this unconventional fermion exhibits Chern number $+4$ at the Fermi level in RhSi (Fig.~\ref{Fig2}(d)).

Our examinations of the unconventional fermions at $R$ with and without SOC (SM C.2) confirm the results of previous analyses of SGs 19 and 198~\cite{Adrian19,BalatskySG19,DoubleDirac, Layer_group, NewFermion}.  When SOC is taken into consideration, RhSi displays at $R$ a six-fold-degenerate chiral double spin-1 Weyl at $\sim 0.4 eV$ below the Fermi level, which at the finite-$q$ gap spanned by the Fermi energy exhibits Chern number $-4$ (Fig.~\ref{Fig2}(b,e)).  Projecting out of our TB model the six-band subspace of this chiral fermion results in a $k\cdot p$ theory related by a unitary transformation to that presented in Ref.~\onlinecite{NewFermion} of two coupled spin-1 fermions with individual $\vec{k}\cdot\vec{S}$ dispersion.

Calculating the surface states of RhSi (Fig.~\ref{Fig3}(a)) through surface Green's functions (SM A), we find that the $(001)$-surface displays four topological Fermi arcs connecting the projections of the bulk chiral fermions at $\bar{\Gamma}$ to those at $\bar{M}$ across the entire surface BZ.  Unlike the recently observed trivial arcs in WTe$_{2}$ ~\cite{trivialarc, LaAlGe}, the long Fermi arcs in RhSi are guaranteed by bulk topology, and should therefore be robust against changes in surface chemical potential and disorder (SM E).  Though the arcs in our calculations demonstrate a particularly elaborate connectivity (Fig.~\ref{Fig3}(c)), a much simpler direct connectivity is also allowed (Fig.~\ref{Fig3}(d)). We also find the Fermi arcs to have $\sim 80\%$ spin polarization~\cite{arcDetect2} (Fig.~\ref{Fig3}(b)).  Therefore, RhSi is also an attractive platform for spintronic applications~\cite{spintronics1,spintronics2}.

To summarize our analysis of the electronic structure of RhSi, we find that it is a remarkably ideal candidate for the observation of chiral transport and optical phenomena and for the direct examination of unconventional fermions.  Bands within the $k\cdot p$ regime of the unconventional fermions at $\Gamma$ and $R$ cleanly characterize the \emph{entire} Fermi surface, such that the separation between Fermi pockets of opposite Chern number is the \emph{entire length} of the 3D diagonal of the BZ cube.  The remaining bulk band manifolds are otherwise separated by a gap of $\sim$ 1.2 eV (Fig.~\ref{Fig1}(d)), such that RhSi has \emph{by far} the largest topologically nontrivial energy window of any previously proposed or experimentally realized chiral semimetal (SM E).  RhSi also therefore displays on its surface topologically-guaranteed Fermi arcs that span \emph{the entire} surface BZ, and uniquely come in time-reversed pairs (Fig.~\ref{Fig3}).  Finally, unlike in previous band-inversion Weyl semimetals where pairs of Weyl points have been related by mirror symmetry, the chiral fermions in RhSi are free to sit with an energy offset, enabling chiral photogalvanic transport~\cite{photogalvanic}.

We therefore conclude with a numerical prediction of quantized optical transport in RhSi.  In Ref.~\onlinecite{photogalvanic}, the authors show that in a structurally chiral system for which only a single two-band Weyl fermion is partially unoccupied, such as a Kramers Weyl metal~\cite{KramersWeyl}, the difference in the rate of current density resulting from exciting electrons with left- and right-handed circularly polarized light is quantized in terms of fundamental constants:
\begin{equation}
\frac{dj}{dt} = \frac{2I\beta_{0}}{c\epsilon_{0}}C,\ \beta_{0} = \frac{\pi e^{3}}{h^2},
\label{eq:photocurrent}
\end{equation}
where $I$ is the intensity of applied light and $C$ is the Chern number of the Weyl point (Fig.~\ref{Fig4}(a,b)).  In RhSi, the four-fold fermion at $\Gamma$ sits just above the Fermi energy while the chiral double spin-1 Weyl at $R$ sits below and is fully occupied; the location of the chiral fermions in its band structure (Fig.~\ref{Fig1}(d)) is \emph{practically identical} to the ideal case proposed in Ref.~\onlinecite{photogalvanic}.  We observe that the angular momentum selection rules for circularly polarized light appear to strongly constrain the allowed transitions in this four-fold fermion, such that only transitions between bands with $\Delta m_{j}=\pm 1$ contribute to the photocurrent~\cite{selectionrules}.  Therefore, when weighting by Fermi occupation factors, the photocurrent rate calculated from the trace of the gyrotropic tensor (SM D), though initially fluctuating, still saturates at the quantized value $(2I\beta_{0}/c\epsilon_{0})\times 4$ with increasing incident photon energy $E_{p}$ in the vicinity of $\Gamma$, or four times the value predicted for a conventional Weyl fermion (Fig.~\ref{Fig4}(b,d)).  Therefore, despite the multiband complexities of its unconventional chiral fermions, RhSi remains a plausible candidate for probing the quantized photogalvanic effect.

\begin{acknowledgements}
The authors thank Charles L.  Kane, Barry Bradlyn Jennifer Cano, and B. A. Bernevig for discussions. The work at Princeton is supported by the National Science Foundation, Division of Materials Research, under Grants No. NSF-DMR-1507585 and No. NSF-DMR-1006492 and by the Gordon and Betty Moore Foundation through the EPIQS program Grant No. GBMF4547-HASAN. The work at the National University of Singapore was supported by the National Research Foundation, Prime Minister's Office, Singapore under its NRF fellowship (NRF Award No. NRF-NRFF2013-03).  B. J. W. was supported through Nordita under ERC DM 321031. The work at Northeastern University was supported by the US Department of Energy (DOE), Office of   Science, Basic Energy Sciences Grant No. DE-FG02-07ER46352, and benefited from Northeastern University's Advanced Scientific Computation Center (ASCC) and the NERSC supercomputing center through DOE Grant No. DE-AC02-05CH11231. The work at the National Sun Yat-sen University was supported by the Ministry of Science and Technology in Taiwan under Grant No. MOST105-2112-M-110-014-MY3. T.-R.C. is supported by the Ministry of Science and Technology and National Cheng Kung University, Taiwan. T.-R.C. also thanks National Center for Theoretical Sciences (NCTS), Taiwan for technical support.
\end{acknowledgements}

\section{Note Added}
We notice a related work that reports similar chiral fermions~\cite{CoSi_zhang}.

\clearpage

\begin{figure}[t]

\includegraphics[width=140mm]{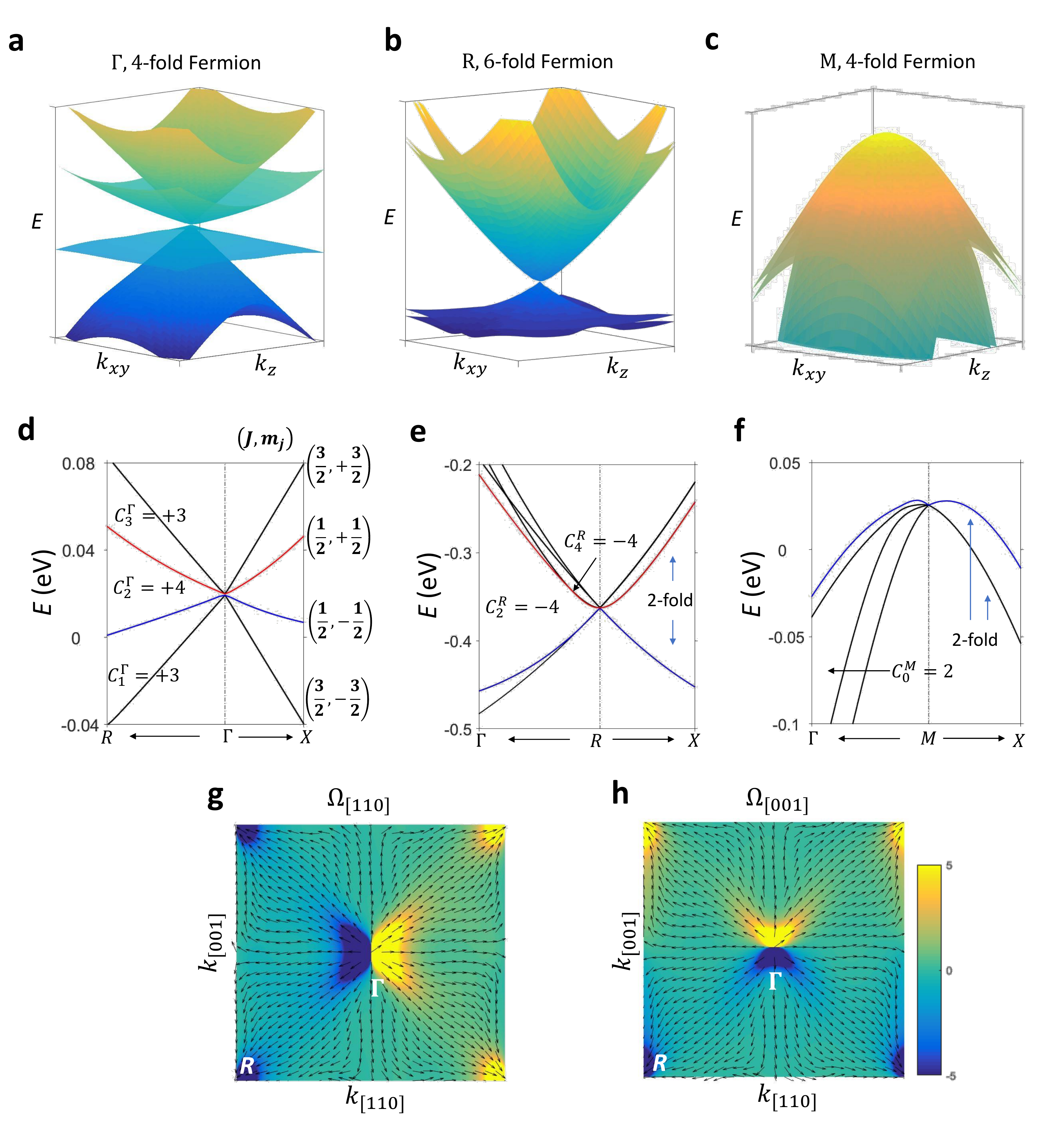}
\caption{\textbf{Energy dispersions and chiral character of the four-fold- and six-fold-degenerate unconventional fermions in RhSi.}  (a,b,c) 3D energy dispersions of the degeneracies at $\Gamma$, $R$, and $M$, respectively.  (d,e,f) Band structures in the vicinities of $\Gamma$, $R$, and $M$, respectively.  Due to the local Kramers theorem enforced under the combined operation of $(s_{2x,y,z}\times\mathcal{T})^{2}=-1$, bands along $k_{x,y,z}=\pi$ are two-fold-degenerate (e,f). The absence of rotoinversion symmetries in SG 198 allows for nodes at TRIMs to have nontrivial Chern numbers; nodes with multiple finite-$q$ gaps can exhibit different Chern numbers occupying bands up to each gap (SM F) }
\label{Fig2}
\end{figure}

\addtocounter{figure}{-1}
\begin{figure*}[t!]
\caption{(d,e,f).  At the Fermi energy, the four-fold-degenerate fermion at $\Gamma$ has Chern number $+4$ and the double spin-1 Weyl at $R$ has Chern number $-4$.  The quadratic four-fold-degenerate crossing at $M$ (f) also exhibits Chern number, but the bands dispersing from it are almost entirely covered by the Fermi energy.  The four-fold-degenerate unconventional fermion at $\Gamma$ (d) can be considered the combination of two $J=1/2$ and two $J=3/2$ states pinned together by time-reversal and screw symmetries.  The analogous angular momentum eigenvalues for each band can then be deduced by observing the band eigenvalues of $C_{3,111}$ and considering the symmetry-allowed term $\vec{k}\cdot\vec{J}$ (SM C.2).  (g,h) The Berry curvature $\vec{\Omega}$ on the $k_{x}=k_{y}$ plane flows almost directly from $\Gamma$ to $R$ with minimal out-of-plane deviations.  Measuring the intensity of the $xy$ (g) and $z$ (h) components of $\vec{\Omega}$, we verify that $\Gamma$ and $R$ exhibit the local vector fields of $C=\pm 4$ hedgehog defects.}
\label{Fig2}
\end{figure*}

\begin{figure}[t]
\includegraphics[width=140mm]{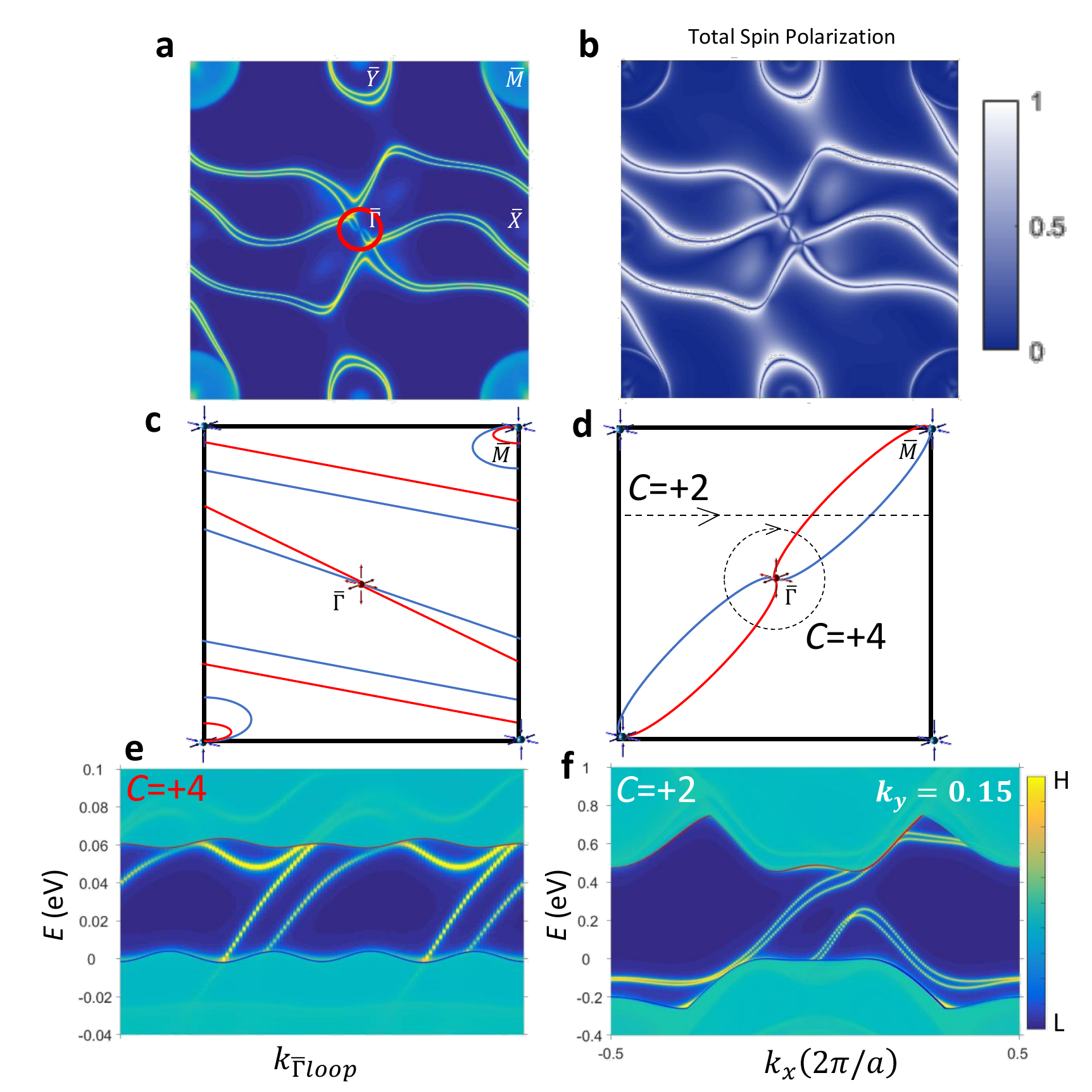}
\caption{ \textbf{Surface state texture of RhSi} (a) The $(001)$-surface states of RhSi calculated using surface Green's functions (SM A).  Four Fermi arcs radiate at $\bar{\Gamma}$ from the projection of the bulk four-fold-degenerate fermion at $\Gamma$, grouping into two time-reversed pairs and spiraling around the BZ (c) until they meet at $\bar{M}$ at the projection of the bulk double spin-1 Weyl at $R$.  (b) The surface states demonstrate $\sim 80\%$ spin polarization (SM A).  (d) An allowed simplified Fermi arc connectivity.  For both possible connectivities (c,d), plotting the surface bands along a clockwise loop surrounding $\bar{\Gamma}$ (red loop in (a), dashed loop in (d)), the surface bands (e) demonstrate a $C=+4$ spectral flow.  (f) Conversely, taking a loop along the zone-spanning dashed line at $k_{y}=0.15$ results in a surface state texture with just $C=+2$ spectral flow, as only two Fermi arcs cross each half of the surface BZ.}
 \label{Fig3}
\end{figure}

\begin{figure}
\includegraphics[width=140mm]{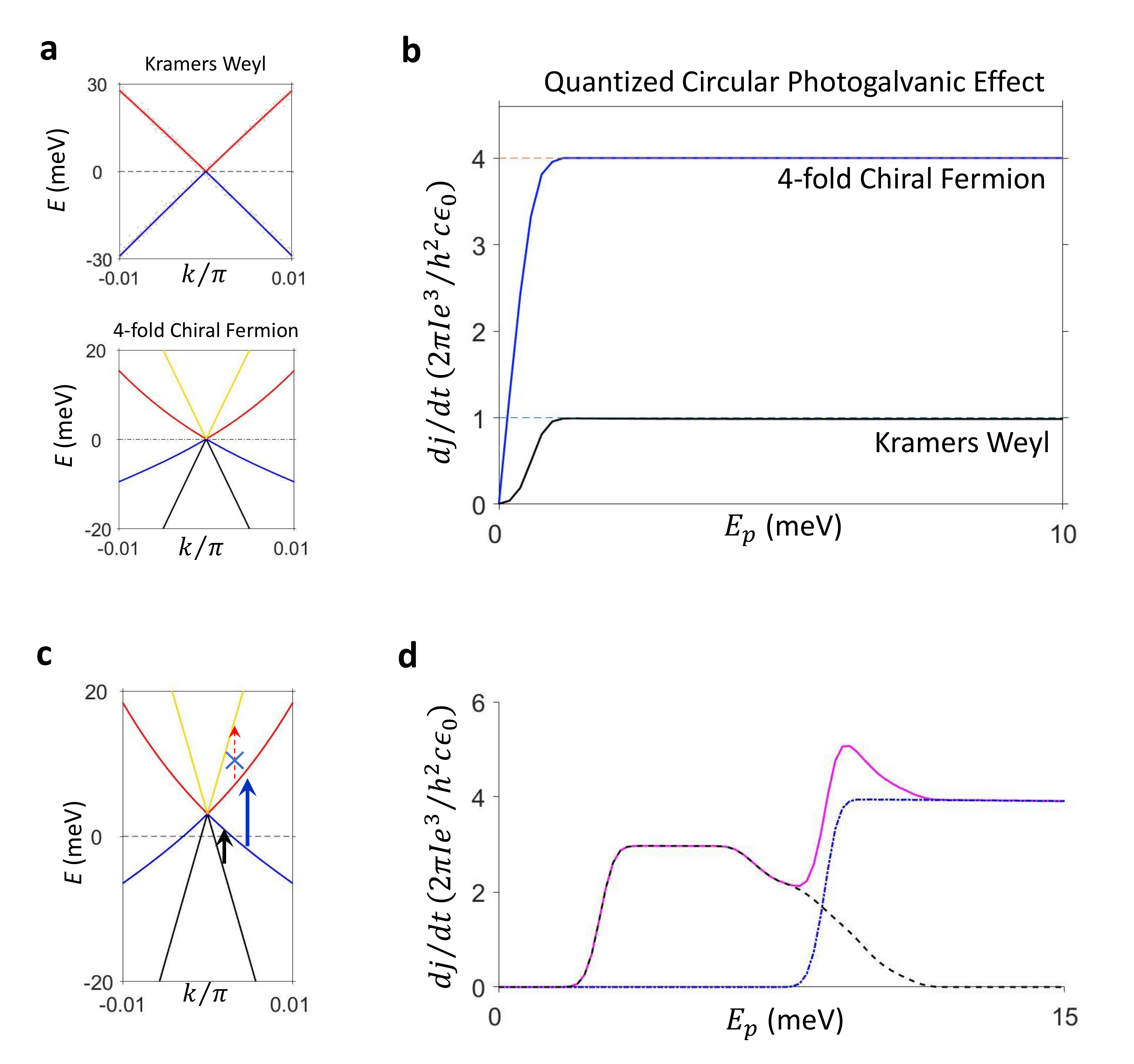}
\caption{\textbf{Quantized circular photogalvanic effect (CPGE) of the four-fold-degenerate unconventional fermion in RhSi} (a,b) The CPGE of a single conventional Weyl node as proposed in Ref.~\cite{photogalvanic}. (b) Calculations of the CPGE for the four-fold unconventional fermion at $\Gamma$ in RhSi, tuned to half-filling.  The photocurrent rate saturates at four times the value it did for the conventional Weyl in (a), as the Chern number in this gap is four times as large.  (c) The more realistic case of a partial occupation of this four-fold fermion; multiple transitions contribute to the photocurrent. (d) Contributions to the traced photocurrent rate from each transition in (c), calculated from the fitted TB model (SM C.3).  Trend lines in (d) are labeled by the color of their contributing transition in (c), with pink representing the overall photocurrent rate.}
\label{Fig4}
\end{figure}

\clearpage

\newpage

\clearpage

\textbf{
\begin{center}
{\large \underline{Supplementary Material}: \\Unconventional Chiral Fermions and Large Topological Fermi Arcs in RhSi}
\end{center}
}

\vspace{0.2cm}

\begin{center}

\end{center}

\vspace{0.25cm}

\textbf{
\begin{center}
{\large This file includes:\\}
\end{center}
}
\vspace{0.45cm}
\textbf{
\begin{tabular}{l l}
\underline{SM A.} & Density Functional Theoretic Methods\\
\underline{SM B.} & Additional Materials Candidate in Space Group 198 \\
\underline{SM C.} & Tight-Binding Model \\
\underline{SM D.} & Photocurrent Calculation Details\\
\underline{SM E.} & Comparison of RhSi With Previous Examples of Weyl Semimetals \\
\underline{SM F.} & Numerical Calculations of the Chiral Charges of High-fold Fermions  \\
\underline{SM G.} & The Effects of Exchange-correlation Pseudopotentials on RhSi \\
\end{tabular}
}

\newpage

\clearpage
\subsection{\large SM A. Density Functional Theoretic Methods}

First-principles calculations were performed within the density functional theory (DFT) framework using the OPENMX package and the full-potential augmented plane-wave method as implemented in the package WIEN2k~\cite{openmx1,openmx2, DFT2,DFT3}. The generalized gradient approximation (GGA)~\cite{DFT4} was used. Experimentally measured lattice constants were used in DFT calculations of material band structures~\cite{CoSi, CoGe, RhSi, RhGe, AlPd, AlPt, GaPt, GaPd, FeSi, BaPtP}. A $\Gamma$-centered $k$-point $10 \times 10 \times 10$ mesh was used and spin-orbit coupling (SOC) was included in self-consistent cycles.

To generate the $(001)$-surface states of RhSi, Wannier functions were generated using the $p$ orbitals of Si and the $d$ orbitals of Rh. The surface states were calculated for a semi-infinite slab by the iterative Green's function method.  The spin polarization was calculated by the method of Ref.~\onlinecite{arcDetect2}, in which the authors calculate the summed expectation values squared of spin $S_{i}$ in three orthogonal directions $i=x,y,z$:
\begin{equation}
P_{S}=\sqrt{\sum_{i=x,y,z}\langle S_{i}\rangle^{2}}\ .
\end{equation}
\begin{figure}[t]
\includegraphics[width=140mm]{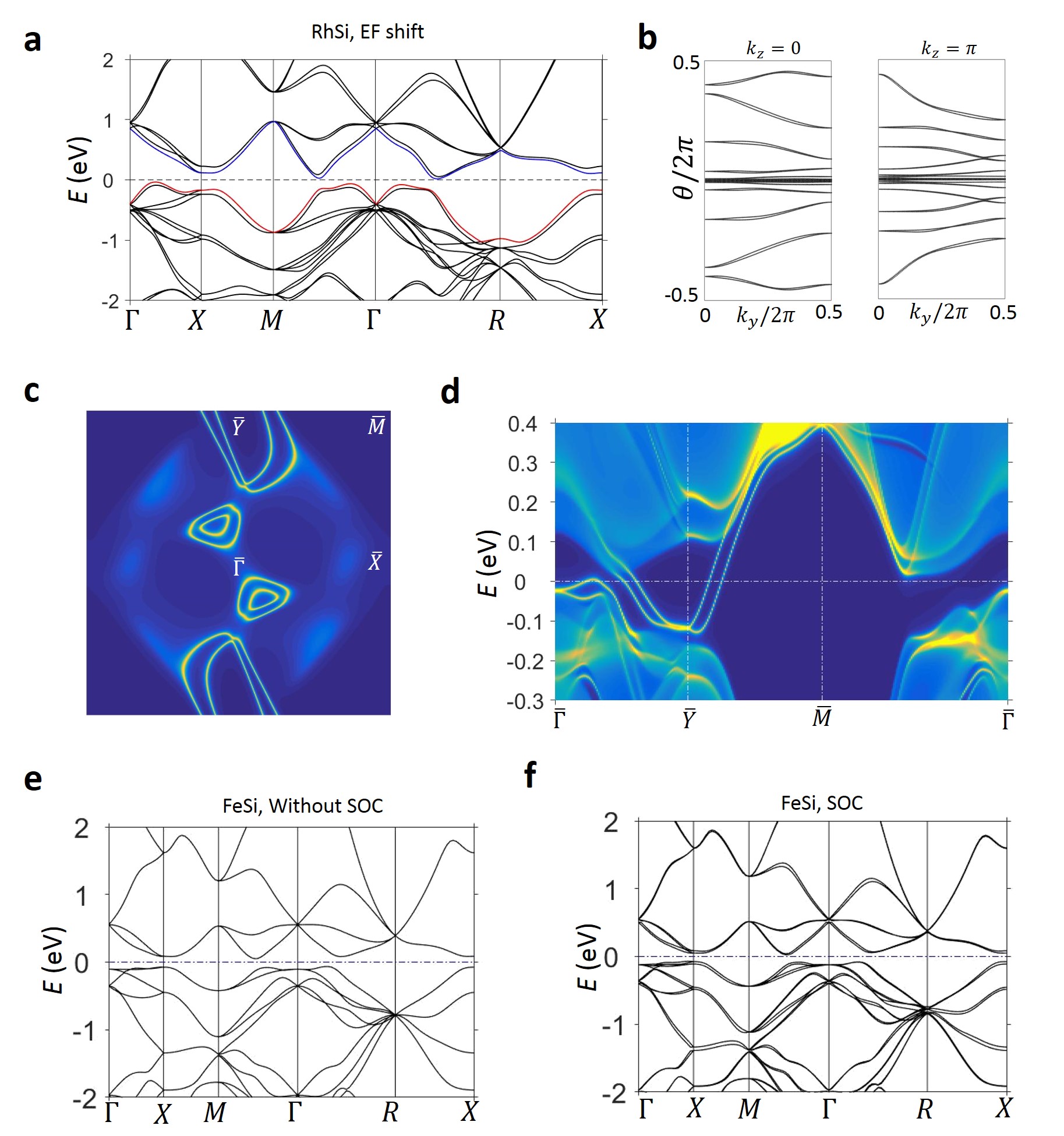}
\caption{\textbf{Trivial gaps in candidate materials} (a) The bands of RhSi after shifting the Fermi level to the gap below. (b) The Wilson loop spectral flow~\cite{WilsonLoop} calculated occupying bands up to this gap indicates that the gap is trivial. (c,d) Surface states calculated by surface Green's functions (SM A) for the new gap in RhSi.  The Fermi arcs close on themselves and have no net spectral flow, further indicating a trivial gap. (e,f) The band structures of FeSi without and with SOC, respectively.  Unlike RhSi, FeSi has a filling of $\nu\in8 \mathbb{Z}$ and, as SG 198 has a minimal insulating filling of $\nu\in 8\mathbb{Z}$~\cite{fillingconstraint}, is a trivial insulator instead of an unconventional fermion semimetal.}
\label{FigS1}
\end{figure}

\subsection{{\large SM B. Additional Materials Candidates in Space Group 198}}

Although unconventional fourfold-degenerate fermions at $\Gamma$ and sixfold-degenerate double spin-1 Weyl points at $R$ are quite common in space group (SG) 198, due to being featured in the minimal band connectivity (SM C.2)~\cite{NewFermion,QuantumChemistry}, not every material featuring them is ideal for experimental applications.  Specifically, both for the measurement of bulk transport effects, such as the quantized photogalvanic effect discussed in SM D, and for the observation of surface Fermi arcs (Fig.~3(a)), a material in SG 198 should have a simple, chiral Fermi surface.  The Fermi surfaces of strong materials candidates should therefore intersect the unconventional fermions at $\Gamma$ and $R$ and be absent of electron and hole pockets.

The first criteria can be achieved by selecting materials with the appropriate electron count.  As noted in Ref.~\onlinecite{fillingconstraint}, SG 198 is cubic supergroup of Bieberbach group SG 19, which has a minimal insulating  filling of $\nu\in 8\mathbb{Z}$.  Furthermore, as shown in Ref.~\onlinecite{QuantumChemistry}, SG 198 has only a single maximal Wyckoff position ($4a$) and band connectivity ($8$), such that bands in SG 198 can never be forced by representation compatibility to form nonminimal connectivities.  Therefore, in the noninteracting limit, crystals in SG 198 with electron counts other than $\nu\in 8\mathbb{Z}$ (such as those, in particular, with $\nu\in 8\mathbb{Z} + 4$) will be metals or semimetals~\cite{Layer_group}.  In Fig.~\ref{FigS1}, we artificially adjust the chemical potential in RhSi as if it had an electron count of $\nu\in 8\mathbb{Z}$, and demonstrate that the resulting Fermi surface is insulating and topologically trivial (Fig.~\ref{FigS1}(a-d)).  As an example of a material in SG 198 that features a trivial gap at the Fermi energy, we present FeSi~\cite{FeSi} (Fig.~\ref{FigS1}(e,f)), which due to its electron count hosts unconventional chiral fermions that fail to contribute to the low-energy physics.

\begin{figure}[t]
\includegraphics[width=140mm]{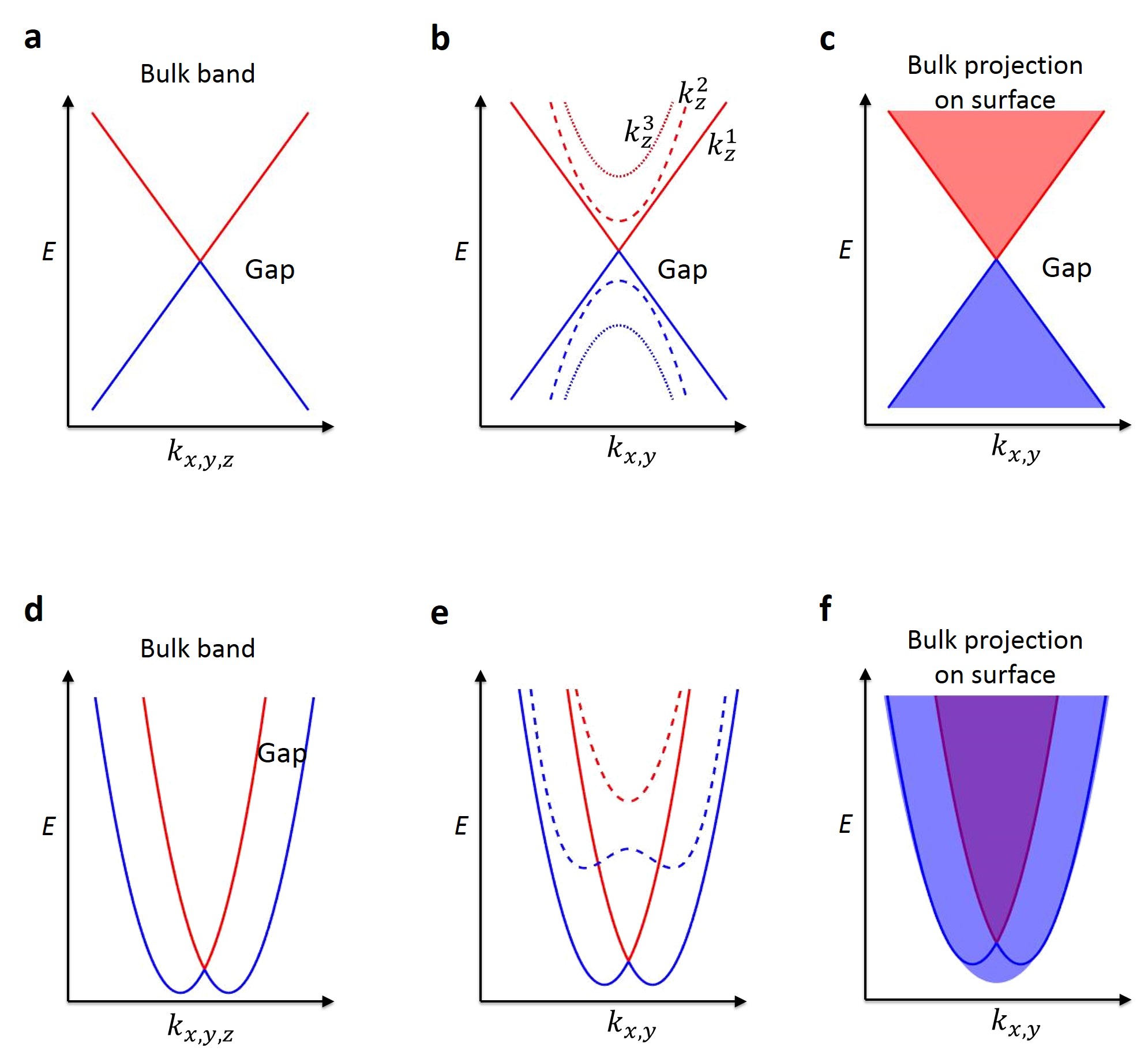}
\caption{\textbf{Chiral fermions and projected surface gaps} (a) Bulk bands near a conventional $|C|=1$ Weyl fermion that isn't tilted. The bands disperse in opposite directions.  (b) Bulk bands for the same Weyl point along the $k_{x}$ or $k_{y}$ directions for different values of $k_{z}$. (c) Surface spectrum of the side surface.  The projected 2D triangles are filled by the projections of the 3D bulk Weyl cones. (d) Bulk bands of a chiral point generated by Rashba-like splitting. (e) Bands in the vicinity of this point for different $k_{z}$ values. (f) On the side surface, the Rashba cones project to fill in all of the surface gaps in the vicinity of the projected chiral point, and therefore obscure any Fermi arc states on this surface.}
\label{FigS2}
\end{figure}

The second criteria, that the Fermi surface be free of electron and hole pockets, is necessary to observe surface Fermi arcs.  More specifically, if there is no projected gap at a given surface, Fermi arcs cannot be resolved.  As discussed in the main text and in more detail in Ref.~\onlinecite{NewFermion}, a fully isolated N-fold-degenerate chiral fermion can have up to N-1 band gaps, each of which can host its own nontrivial Chern number.  However, not all gaps correspond to observable surface states; in many cases, the projections of bulk bands cover possible surface gaps and obscure topological Fermi arcs.  Specifically, when bands disperse from a chiral fermion in opposite directions, such as in a conventional $|C|=1$ Weyl that isn't tilted, a projected bulk gap is allowed (Fig.~\ref{FigS2}(a-c)).  However, in cases like bands split by Rashba coupling (Fig.~\ref{FigS2}(d-f)), all surface gaps will be filled by projected bulk bands and Fermi arcs will not be visible.
\begin{figure}[t]
\includegraphics[width=140mm]{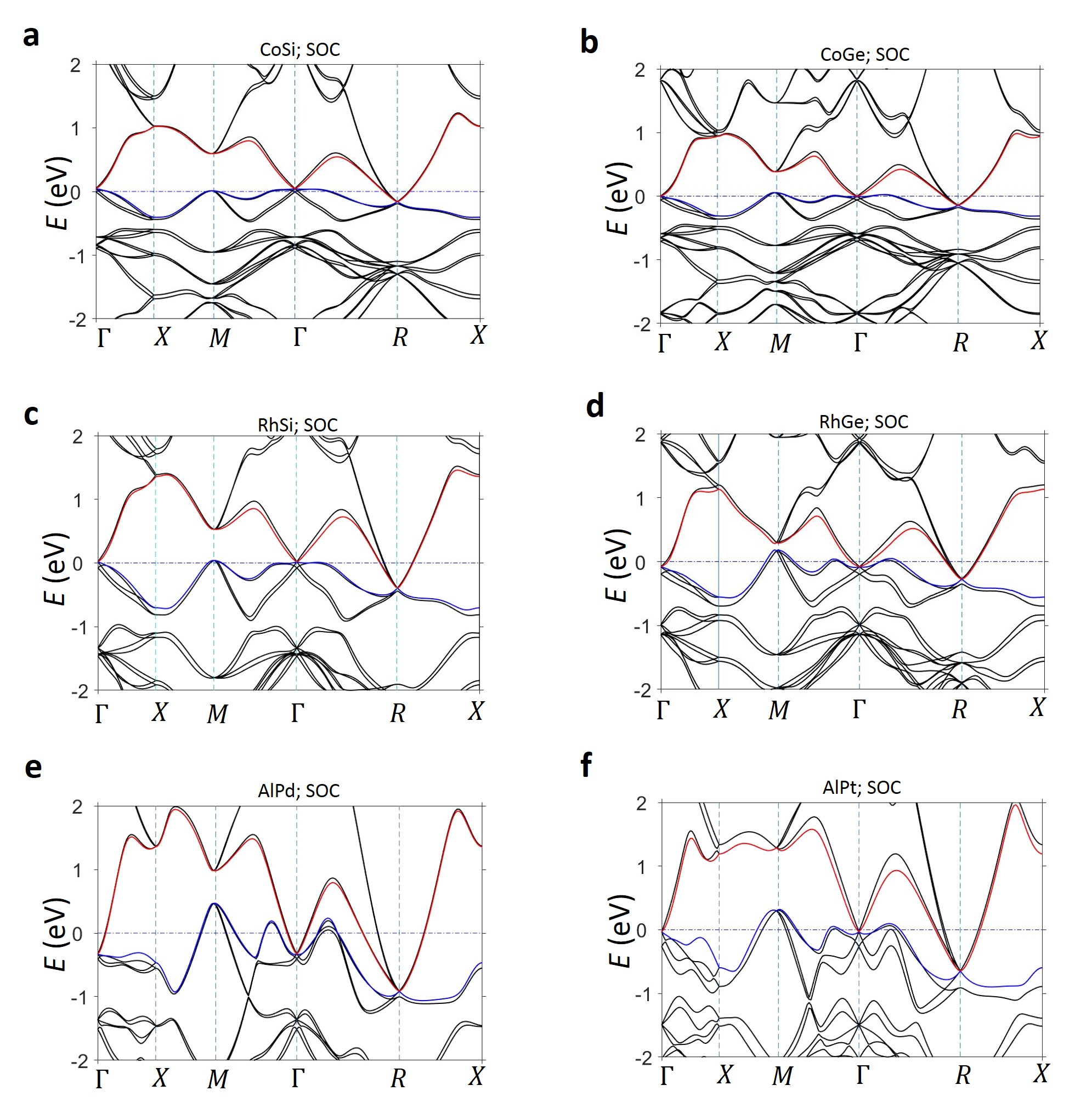}
\caption{\textbf{Band structures of promising material candidates in the RhSi or isostructural materials families in SG 198} (a-f) SOC bands calculated by DFT for CoSi, CoGe, RhSi, RhGe, AlPd, and AlPt, respectively.  CoSi, AlPd, and AlPt were previously identified as hosting sixfold-degenerate unconventional fermions in Ref.~\onlinecite{NewFermion}.}
\label{FigS3}
\end{figure}

We have identified several materials candidates in SG 198 that satisfy the above conditions for experimental viability.  In the main text, we highlight RhSi~\cite{RhSi}, which we found in our materials search to be the strongest candidate for observing large Fermi arcs and bulk photogalvanic effects (SM D).  There are  also several other promising materials candidates isostructural to RhSi, which include CoSi~\cite{CoSi}, CoGe~\cite{CoGe}, RhGe~\cite{RhGe}, AlPd~\cite{AlPd}, AlPt~\cite{AlPt}, GaPd~\cite{GaPd}, and GaPt~\cite{GaPt}.  Of these materials, CoSi, AlPd, and AlPt were previously identified as hosting sixfold-degenerate unconventional fermions in Ref.~\onlinecite{NewFermion}. In Fig.~\ref{FigS2}, we plot band structures calculated through DFT for CoSi (a), CoGe (b), RhSi (c), RhGe (d), AlPd (e), and AlPt (f).  We additionally identified promising materials candidates in the BaPtX (X=P,As,Sb) family of materials~\cite{BaPtP}, for which the unit cell structure and bands calculated through DFT are plotted in Fig.~\ref{FigS3}.
\begin{figure}[t]
\includegraphics[width=140mm]{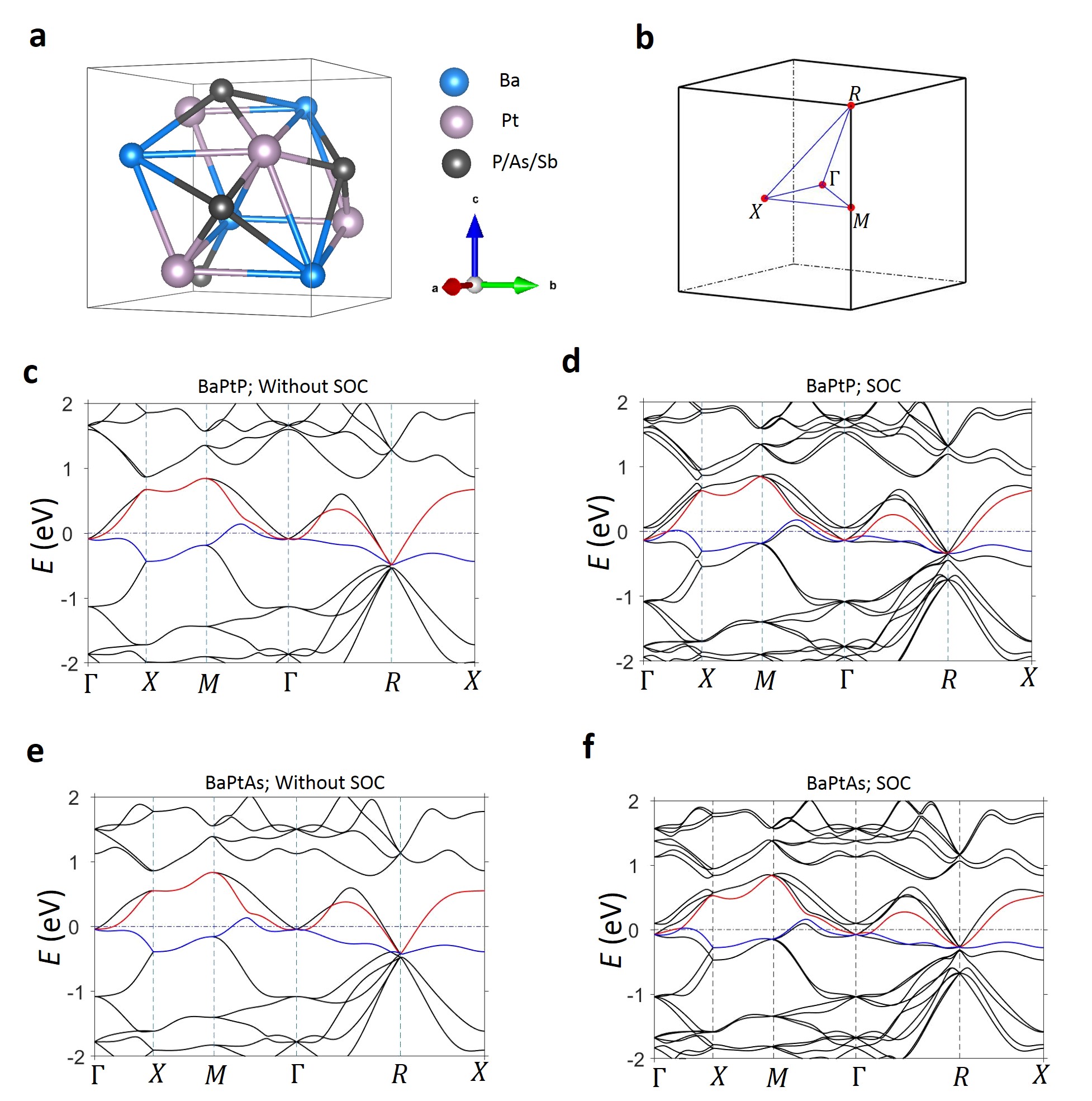}
\caption{\textbf{Band structures of promising material candidates in the BaPtAs family in SG 198} (a) The crystal structure of BaPtX (X=P, As, Sb). (b) Cubic Brillouin zone of SG 198.  (c,d) The band structures of BaPtP without and with SOC. (e,f) The band structures of BaPtAs without and with SOC.}
\label{FigS4}
\end{figure}

\subsection{\large SM C. Tight-Binding Model}

\subsubsection{\large SM C.1 Symmetry-Generation of Hopping Terms}

To construct a tight-binding model for SG 198, we begin by choosing a four-site unit cell for which each site carries only an $s$ orbital.  Confirming the results of Ref.~\onlinecite{QuantumChemistry} for SG 198, we find that when SOC is on, the minimal band connectivity required to capture the insulating filling constraint $\nu\in 8\mathbb{Z}$~\cite{fillingconstraint} is indeed sufficiently captured by placing $s_{1/2}$ orbitals at Wyckoff positions with the minimum multiplicity of $4$.  Starting with a site designated $A$ at unit cell position $(0,0,0)$ in units of the orthogonal lattice constants $a_{x,y,z}$, we utilize an allowed permutation of the group generators as listed in Bradley and Cracknell~\cite{BigBook}:
\begin{eqnarray}
s_{2x}=\left\{C_{2x}\bigg|\frac{1}{2}\frac{1}{2}0\right\},\ &s_{2y}&=\left\{C_{2y}\bigg|0\frac{1}{2}\frac{1}{2}\right\}\nonumber \\
s_{2z}=\left\{C_{2z}\bigg|\frac{1}{2}0\frac{1}{2}\right\},\ &C_{3,111}&=\left\{C_{3,111}\bigg|000\right\}.
\label{gen198}
\end{eqnarray}

These generators, three nonintersecting screws related by cubic threefold rotation, generate three additional sites in the unit cell: $B$ located at $(\frac{1}{2},\frac{1}{2},0)$, $C$ located at $(\frac{1}{2},0,\frac{1}{2})$, and $D$ located at $(0,\frac{1}{2},\frac{1}{2})$.  Graphically, the four sites comprising this model sit at the four alternating corners of a cube.  A crucial realization here is that there are two distinct $C_{3}$ axes; each site is internally $C_{3}$-symmetric but is cyclically exchanged with two other sites by the $C_{3}$ of the fourth site.  More concretely, under $C_{3,111}$ for example, the $A$ site is left invariant while the other three sites are cyclically rotated $B\rightarrow D\rightarrow C \rightarrow B$. From the $k\cdot p$ results in the next section (SM C.2), we find that this relationship between this cyclic operation and the minimum four-site construction of SG 198 is in fact the reason why its minimum band connectivity displays sixfold-degenerate unconventional fermions.

To construct a concrete model, we employ the method of Ref.~\onlinecite{Layer_group}.  We construct the representations of the group generators at $\Gamma$ which, combined with the mappings they perform on the crystal momenta $\vec{k}$, allow us to enumerate all symmetry-allowed hopping terms up to a specified range in position space.  Written in the mixed notation of the form of the unitary operation that transforms the sublattices Kronecker multiplied by the result of the  $k$-space mapping under the operation, the three screw generators and spinful time-reversal $\mathcal{T}$ take the form:
\begin{eqnarray}
s_{2x}&=&i\tau^{x}\sigma^{x}\otimes (k_{y,z}\rightarrow - k_{y,z}) \nonumber \\
s_{2z}&=& i\mu^{x}\sigma^{z}\otimes (k_{x,y}\rightarrow - k_{x,y}) \nonumber \\
\tilde{\mathcal{T}} &=& i\sigma^{y}K \otimes (k_{x,y,z}\rightarrow -k_{x,y,z})
\end{eqnarray}
where $\tau^{x}$ is $s$-orbital-like hopping between the $A$ and $B$ and the $C$ and $D$ sites, $\mu^{x}$ is $s$-orbital-like hopping between the $A$ and $C$ and $B$ and $D$ sites, and $\sigma$ is the on-site $s_{1/2}$ spinorbital.  The more complicated threefold diagonal rotation $C_{3,111}$ takes the form of a cyclic operation that enforces the mappings:
\begin{eqnarray}
k_{x}&\rightarrow& k_{y} \rightarrow k_{z}\rightarrow k_{x} \nonumber \\
\sigma^{x}&\rightarrow&\sigma^{y}\rightarrow\sigma^{z}\rightarrow\sigma^{x} \nonumber \\
\tau^{x}&\rightarrow&\tau^{x}\mu^{x}\rightarrow\mu^{x}\rightarrow\tau^{x} \nonumber \\
\tau^{y}&\rightarrow&\tau^{x}\mu^{y}\rightarrow\tau^{z}\mu^{y}\rightarrow\tau^{y} \nonumber \\
\tau^{y}\mu^{z}&\rightarrow&\tau^{y}\mu^{x}\rightarrow\mu^{y}\rightarrow\tau^{y}\mu^{z} \nonumber \\
\tau^{x}\mu^{z}&\rightarrow& -\tau^{y}\mu^{y}\rightarrow\tau^{z}\mu^{x}\rightarrow\tau^{x}\mu^{z}
\label{eq:cyclic}
\end{eqnarray}
where in this notation $\tau^{i}\mu^{j}\sigma^{k}$ stands for the Kronecker product  $s^{i}\otimes s^{j} \otimes s^{j}$ of Pauli matrices $s^{x,y,z}$ and the $2\times 2$ identity $s^{0}$, and the identity terms $s^{0}$ are suppressed.

We enumerate symmetry-allowed hopping terms up to first-nearest-neighbor interactions, finding that there are $2$ SOC-free terms and $6$ terms with SOC.  The overall tight-binding Hamiltonian $\mathcal{H}_{198}(\vec{k})$ can be expressed as the sum of its spinless part $\tilde{\mathcal{H}}_{198}(\vec{k})$ and the spin-orbit terms $V_{s/r,i}(\vec{k})$:
\begin{equation}
\mathcal{H}_{198}(\vec{k}) = \tilde{\mathcal{H}}(\vec{k})_{198} + \sum_{i=1,2,3}V_{r,i}(\vec{k}) + V_{s,i}(\vec{k})
\label{eq:tb}
\end{equation}
where
\begin{eqnarray}
\tilde{\mathcal{H}}_{198}(\vec{k}) &=& v_{1}\bigg[\tau^{x}\cos\left(\frac{k_{x}}{2}\right)\cos\left(\frac{k_{y}}{2}\right) + \tau^{x}\mu^{x}\cos\left(\frac{k_{y}}{2}\right)\cos\left(\frac{k_{z}}{2}\right) + \mu^{x}\cos\left(\frac{k_{z}}{2}\right)\cos\left(\frac{k_{x}}{2}\right)\bigg] \nonumber \\
&+&  v_{p}\bigg[\tau^{y}\mu^{z}\cos\left(\frac{k_{x}}{2}\right)\sin\left(\frac{k_{y}}{2}\right) + \tau^{y}\mu^{x}\cos\left(\frac{k_{y}}{2}\right)\sin\left(\frac{k_{z}}{2}\right) + \mu^{y}\cos\left(\frac{k_{z}}{2}\right)\sin\left(\frac{k_{x}}{2}\right)\bigg] \nonumber \\
\end{eqnarray}
and
\begin{eqnarray}
V_{r1}(\vec{k})&=&v_{r1}\bigg[\tau^{y}\mu^{z}\sigma^{y}\cos\left(\frac{k_{x}}{2}\right)\cos\left(\frac{k_{y}}{2}\right) + \tau^{y}\mu^{x}\sigma^{z}\cos\left(\frac{k_{y}}{2}\right)\cos\left(\frac{k_{z}}{2}\right) + \mu^{y}\sigma^{x}\cos\left(\frac{k_{z}}{2}\right)\cos\left(\frac{k_{x}}{2}\right)\bigg] \nonumber \\
V_{r2}(\vec{k})&=&v_{r2}\bigg[\tau^{y}\sigma^{z}\cos\left(\frac{k_{x}}{2}\right)\cos\left(\frac{k_{y}}{2}\right) + \tau^{x}\mu^{y}\sigma^{x}\cos\left(\frac{k_{y}}{2}\right)\cos\left(\frac{k_{z}}{2}\right) + \tau^{z}\mu^{y}\sigma^{y}\cos\left(\frac{k_{z}}{2}\right)\cos\left(\frac{k_{x}}{2}\right)\bigg] \nonumber \\
V_{r3}(\vec{k})&=&v_{r3}\bigg[\tau^{y}\mu^{z}\sigma^{x}\sin\left(\frac{k_{x}}{2}\right)\sin\left(\frac{k_{y}}{2}\right) + \tau^{y}\mu^{x}\sigma^{y}\sin\left(\frac{k_{y}}{2}\right)\sin\left(\frac{k_{z}}{2}\right) + \mu^{y}\sigma^{z}\sin\left(\frac{k_{z}}{2}\right)\sin\left(\frac{k_{x}}{2}\right)\bigg] \nonumber \\
V_{s1}(\vec{k})&=&v_{s1}\bigg[\tau^{x}\sigma^{x}\sin\left(\frac{k_{x}}{2}\right)\cos\left(\frac{k_{y}}{2}\right) + \tau^{x}\mu^{x}\sigma^{y}\sin\left(\frac{k_{y}}{2}\right)\cos\left(\frac{k_{z}}{2}\right) + \mu^{x}\sigma^{z}\sin\left(\frac{k_{z}}{2}\right)\cos\left(\frac{k_{x}}{2}\right)\bigg] \nonumber \\
V_{s2}(\vec{k})&=&v_{s2}\bigg[\tau^{x}\sigma^{y}\cos\left(\frac{k_{x}}{2}\right)\sin\left(\frac{k_{y}}{2}\right) + \tau^{x}\mu^{x}\sigma^{z}\cos\left(\frac{k_{y}}{2}\right)\sin\left(\frac{k_{z}}{2}\right) + \mu^{x}\sigma^{x}\cos\left(\frac{k_{z}}{2}\right)\sin\left(\frac{k_{x}}{2}\right)\bigg] \nonumber \\
V_{s3}(\vec{k})&=& v_{s3}\bigg[\tau^{x}\mu^{z}\sigma^{z}\cos\left(\frac{k_{x}}{2}\right)\sin\left(\frac{k_{y}}{2}\right) - \tau^{y}\mu^{y}\sigma^{x}\cos\left(\frac{k_{y}}{2}\right)\sin\left(\frac{k_{z}}{2}\right) + \tau^{z}\mu^{x}\sigma^{y}\cos\left(\frac{k_{z}}{2}\right)\sin\left(\frac{k_{x}}{2}\right)\bigg]. \nonumber \\
\end{eqnarray}

\subsubsection{\large SM C.2 $k\cdot p$ Theory of $R$ and $\Gamma$ Without and With SOC}

To develop $k\cdot p$ models of the unconventional fermions at $R$ and $\Gamma$, we  begin by keeping the full eight-band matrix structure at each TRIM from the tight-binding model in SM C.1 and note when algebra allows for higher degeneracy.   For threefold- and sixfold-degenerate fermions, this approach is superior to projecting out three- and six-dimensional subspaces, as doing so removes our ability to exploit the Clifford algebra of $8\times 8$ matrices to prove that no other gaps may be opened at  either TRIM.  For the fourfold-degenerate unconventional fermion at $\Gamma$, we additionally exploit the fact that the Little group at $\Gamma$ is isomorphic to chiral point group $23$ ($T$) and use an understanding of the corepresentations of this point group to form a four-band $k\cdot p$ theory~\cite{KramersWeyl,BigBook}. Therefore, procedurally, we begin by expanding the tight-binding model around each TRIM.   We then use the eight-band  forms of the group generators at  the TRIMs to exhaustively check whether any additional mass terms are allowed.  As we show below, all of the gaps at $\Gamma$ and $R$ are fully opened just by considering the symmetry-allowed first-nearest-neighbor hopping terms from SM C.1.

We begin at $R$ ($\vec{k}=(\pi,\pi,\pi)$).  When SOC is weak, the system gains an additional $SU(2)$ invariance and a second set of spinless symmetries are additionally enforced, which we denote with tildes.  At $R$ under weak SOC, the spinless screws and time-reversal take the forms:
\begin{equation}
\tilde{s}_{2x}=i\tau^{y}\mu^{z},\ \tilde{s}_{2z}=i\mu^{y},\ \tilde{\mathcal{T}}=K,
\end{equation}
the screws mutually anticommute:
\begin{equation}
\{\tilde{s}_{2i},\tilde{s}_{2j}\}=-2\delta_{ij},
\end{equation}
and the spinless threefold rotation $\tilde{C}_{3,111}$ exchanges the screws by the same cyclic relations listed above.  As the screws and $\tilde{\mathcal{T}}$, the generators of SG 19~\cite{BigBook}, themselves enforce a $4\times 2$-fold degeneracy through this algebra~\cite{Adrian19,BalatskySG19,Layer_group}, our eight-band model of SG 198 also displays a spinless  eightfold-degenerate chiral fermion~\cite{DoubleDirac,BalatskySG19} at $R$ under vanishing SOC:
\begin{equation}
\tilde{\mathcal{H}}_{198,R} = v_{R}\left(\tau^{y}\mu^{z}k_{x} + \tau^{y}\mu^{x}k_{y} + \mu^{y}k_{z}\right)
\end{equation}
where minus signs have been suppressed in expanding $\tilde{\mathcal{H}}_{198}(R)$.

Introducing SOC, we break $SU(2)$ invariance and are left at $R$ with spinful screws and time-reversal:
\begin{equation}
s_{2i}=\tilde{s}_{2i}\otimes i\sigma^{i},\ \mathcal{T}=i\sigma^{y}K.
\end{equation}
The spinful threefold rotation $C_{3,111}$ is the same as the spinless operation $\tilde{C}_{3,111}$ times the additional cyclic rotation of the Pauli matrices described in Eq.~(\ref{eq:cyclic}).  These spinful screws now commute instead of anticommute:
\begin{equation}
[s_{2i},s_{2j}]=0
\end{equation}
and all square to $+1$.  Under these conditions, our minimal eight-band model admits just a single mass term at $R$:
\begin{equation}
V_{m,R} = m_{s,R}\left[\tau^{y}\mu^{z}\sigma^{x} + \tau^{y}\mu^{x}\sigma^{y} + \mu^{y}\sigma^{z}\right]
\end{equation}
which derives from $V_{r,3}(R)$.  Though many linear terms are allowed near $R$, we can use our tight-binding model to constrain the choices to those originating from first-nearest-neighbor hopping:
\begin{eqnarray}
\mathcal{H}_{198,R}(\vec{k}) &=& V_{m,R} + V_{v,R}(\vec{k}) \nonumber \\
V_{v,R}(\vec{k})&=& v_{s1,R}\left[\mu^{x}\sigma^{z}k_{x} + \tau^{x}\sigma^{x}k_{y} + \tau^{x}\mu^{x}\sigma^{y}k_{z}\right] \nonumber \\
&+& v_{s2,R}\left[\tau^{x}\sigma^{y}k_{x} + \tau^{x}\mu^{x}\sigma^{z}k_{y} + \mu^{x}\sigma^{x}k_{z}\right] \nonumber \\
&+& v_{s3,R}\left[\tau^{x}\mu^{z}\sigma^{z}k_{x} - \tau^{y}\mu^{y}\sigma^{x}k_{y} + \tau^{z}\mu^{x}\sigma^{y}k_{z}\right].
\end{eqnarray}
This $k\cdot p$ Hamiltonian describes a sixfold-degenerate chiral double spin-1 Weyl point and a twofold-degenerate quadratic band.  These two band features can be considered as the minimal degeneracy required by the symmetry algebra (twofold degeneracy by Kramers theorem) and a higher-fold degeneracy \emph{permitted} to be gapless by the algebra.  Specifically, if one represents the spinless parts of the screws and threefold rotation using $3\times 3$ matrices in terms of coordinate transformations (or as if they acted on $p$ orbitals~\cite{QuantumChemistry}):
\begin{eqnarray}
\tilde{C}_{3,111} = \left(\begin{array}{ccc}
0 & 1 & 0 \\
0 & 0 & 1 \\
1 & 0 & 0
\end{array}\right),\ \tilde{s}_{2x}= \left(\begin{array}{ccc}
1 & 0 & 0 \\
0 & -1 & 0 \\
0 & 0 & -1
\end{array}\right), \tilde{s}_{2z}= \left(\begin{array}{ccc}
-1 & 0 & 0 \\
0 & -1 & 0 \\
0 & 0 & 1
\end{array}\right),
\label{eq:p3}
\end{eqnarray}
one finds that a threefold degeneracy is already required.  The addition of spinful time-reversal $\mathcal{T}$ doubles this degeneracy, resulting in the algebraic description of a double spin-1 Weyl point.   This confirms the results of Ref.~\onlinecite{NewFermion}, in which the authors conclude that the sixfold-degenerate spinful chiral fermion at $R$ in SG 198 can be described by coupling two threefold-degenerate fermions that individually have Hamiltonians of the form $\vec{k}\cdot\vec{S}$.

At $\Gamma$, the picture without SOC is nearly identical to that at $R$ with SOC.  The spinless screws are cyclically exchanged under spinless threefold rotation $\tilde{C}_{3,111}$, mutually commute:
\begin{equation}
[\tilde{s}_{2i},\tilde{s}_{2j}]=0,
\end{equation}
and square to $+1$.  This algebra can be satisfied with a scalar (a $1\times 1$ matrix), or with $3\times 3$ matrices of the form of Eq.~(\ref{eq:p3}), locally protecting $1\times 2$- and $3\times 2$-fold degeneracies, respectively.  We find that our minimal eight-band model with weak SOC permits just a single mass term at $\Gamma$ such that the spinless $k\cdot p$ Hamiltonian takes the form:
\begin{eqnarray}
\tilde{\mathcal{H}}_{198,\Gamma}(\vec{k}) &=& m_{\Gamma}\left[\tau^{x} + \tau^{x}\mu^{x} + \mu^{x}\right] \nonumber \\
&+& v_{\Gamma}\left[\mu^{y}k_{x} + \tau^{y}\mu^{z}k_{y} + \tau^{y}\mu^{x}k_{z}\right]
\end{eqnarray}
where $m_{\Gamma}$ originates from $v_{1}(\Gamma)$.  The band structure, analogously to that $R$ under strong SOC, features a twofold-degenerate quadratic band and a spinless double spin-1 Weyl point~\cite{NewFermion}.

When SOC is introduced at $\Gamma$, the band features and gaps are significantly altered.  The twofold-degenerate quadratic band opens up into a Kramers Weyl fermion~\cite{KramersWeyl} and two more mass terms are allowed:
\begin{eqnarray}
V_{m,\Gamma} &=&m_{s1,\Gamma}\left[\tau^{y}\mu^{z}\sigma^{y} + \tau^{y}\mu^{x}\sigma^{z} + \mu^{y}\sigma^{x}\right] \nonumber \\
&+&  m_{s2,\Gamma}\left[\tau^{y}\sigma^{z} + \tau^{x}\mu^{y}\sigma^{x} + \tau^{z}\mu^{y}\sigma^{y}\right],
\end{eqnarray}
which originate from $V_{r1}(\Gamma)$ and $V_{r2}(\Gamma)$, respectively.  Again choosing linear terms originating from the first-nearest-neighbor tight-binding model expanded around $\Gamma$, we develop a $k\cdot p$ Hamiltonian for all eight bands:
\begin{equation}
\mathcal{H}_{198,\Gamma}(\vec{k}) = \tilde{\mathcal{H}}_{198,\Gamma}(\vec{k}) + V_{m,\Gamma} + V_{v,\Gamma}(\vec{k}),
\label{eq:gammakp}
\end{equation}
where
\begin{eqnarray}
V_{v,\Gamma}(\vec{k})&=&v_{s1,\Gamma}\left[\tau^{x}\sigma^{x}k_{x} + \tau^{x}\mu^{x}\sigma^{y}k_{y} + \mu^{x}\sigma^{z}k_{z}\right] \nonumber \\
&+& v_{s2,\Gamma}\left[\mu^{x}\sigma^{x}k_{x} + \tau^{x}\sigma^{y}k_{y} + \tau^{x}\mu^{x}\sigma^{z}k_{z}\right] \nonumber \\
&+& v_{s3,\Gamma}\left[\tau^{z}\mu^{x}\sigma^{y}k_{x} + \tau^{x}\mu^{z}\sigma^{z}k_{y} - \tau^{y}\mu^{y}\sigma^{x}k_{z}\right]. \nonumber \\
\end{eqnarray}
This $k\cdot p$ Hamiltonian describes two, twofold-degenerate Kramers Weyl fermions~\cite{KramersWeyl} split from a previously uncharacterized fourfold-degenerate chiral unconventional fermion.  The Kramers Weyl points represent the minimum twofold degeneracy necessary to support the generator algebra at this point~\cite{Layer_group,KramersWeyl}, and the fourfold degeneracy represents a subtle gapless $4\times 4$ algebra which is also permitted.  To understand this fourfold degeneracy, we make the analogy to first describing the weak-SOC spinless double spin-1 Weyl point at $\Gamma$ by three $p$ orbitals and an uncoupled electron spin.  We choose the $p$ orbitals to be oriented along the $111$ direction, and as such define new axes $z'$ along the $111$ direction and $x'$ and $y'$ spanning the plane normal to $z'$.  Expressing the three $p$ orbitals as linear combinations with distinct angular momentum eigenvalues $(l,m_{l})$, we arrive at an $l=0,m_{l}=0,\ p_{z'}$ orbital and $l=1,m_{l}=\pm 1,\ p_{x'}\pm ip_{y'}$ orbitals.  When coupled to the electron spin, which we can choose to also be oriented along $z'$, only total angular momentum $J$ and its components along the $z'$ direction $m_{j}$ are good quantum numbers.  We arrive instead at only a quartet of states related by the spinful symmetry generators: $J=3/2,m_{j}=\pm 3/2$ and $J=1/2,m_{j}=\pm 1/2$ related within pairs by $\mathcal{T}$ and related between pairs by $s_{2x}$.  As it is along the $z'$ direction, $C_{3,111}$ leaves the states invariant and its eigenvalues can be used to deduce $J$.  We can choose a basis in which $s_{2x}$ only acts on the electron spin component in the $z'$ direction $m_{s}$; under this choice $C_{3,111}$ takes the form of its spinless eigenvalues times a spin rotation about the $z'=111$ axis and the screws just act in spin  space.  Up to a sign, the rotationally-invariant linear term $\vec{k}\cdot\vec{J}$ splits the states by $m_{j}$ and can be used to label bands in the $111$ direction as  pictured in Fig.~2(d).

More concretely, we can exploit the methods of Refs.~\onlinecite{DoubleDirac,NewFermion,KramersWeyl} and utilize the group-theoretic description of this point in Ref.~\onlinecite{BigBook} to form an explicit four-band gapless theory of this previously undescribed fourfold-degenerate fermion.  Utilizing the crystalline symmetry textbook Bradley and Cracknell (Ref.~\onlinecite{BigBook}) we identify the fourfold-degenerate fermion as a corepresentation of two-dimensional double-valued irreducible representations $5$ and $6$ of character $G^{9}_{24}$.  As the Little group of the $\Gamma$ point in a crystal is unaffected by projections of lattice translations, it is isomorphic to a point group~\cite{KramersWeyl}.  Here, we find that the Little group at $\Gamma$ is isomorphic to point group $23$ ($T$), and that correspondingly, the fourfold-degenerate fermion is described by the corepresentation of the two-dimensional irreducible representations $^{1}\bar{F}_{23}$ and $^{2}\bar{F}_{23}$.  Using the character table for this point group in Ref.~\onlinecite{BigBook}, we can form representations of the point group generators in a four-band space:
\begin{equation}
s_{2i}=i\sigma^{i},\ C_{3,111}=-(i\tau^{y})^{2/3}(i\sigma^{111})^{2/3},\ \mathcal{T}=i\sigma^{y}K,
\end{equation}
where
\begin{equation}
\sigma^{111}=\frac{1}{\sqrt{3}}\left(\sigma^{x}+\sigma^{y}+\sigma^{z}\right).
\end{equation}
In this basis, we find that to quadratic order, the most general symmetry-allowed Hamiltonian takes the form:
\begin{equation}
\mathcal{H}_{4,\Gamma}(\vec{k})=v_{q}\left(k_{x}^{2} + k_{y}^{2} + k_{z}^{2}\right)\mathds{1}_{\tau\sigma} + v_{1}\big(k_{x}+k_{y}+k_{z}\big)\tau^{y} + v_{2}\big(k_{x}\sigma^{x} + k_{y}\sigma^{y} + k_{z}\sigma^{z}\big),
\end{equation}
where $\mathds{1}_{\tau\sigma}$ is the $4\times4$ identity and  $\mathcal{H}_{4,\Gamma}(\vec{k})$ has the dispersion relation
\begin{equation}
E(\vec{k}) = v_{q}|\vec{k}|^{2} \pm v_{1}\sqrt{3}k_{111}\pm v_{2}|\vec{k}|,
\end{equation}
where
\begin{equation}
k_{111} = \frac{1}{\sqrt{3}}\left(k_{x} + k_{y} + k_{z}\right).
\end{equation}
We can make the association that $\tau^{y}$ indexes orbital angular momentum in the $111$-direction $m_{l}$ and that $\sigma$ represents electron spin $\vec{S}$, under which we recapture our earlier orbital-based description of the fourfold-degenerate fermion.  When the splitting by $m_{l}$ is tuned to be twice the splitting by $m_{s}$,
\begin{equation}
v_{1}=2v_{2},
\end{equation}
the Hamiltonian $\mathcal{H}_{4,\Gamma}(\vec{k})$ takes the form $\vec{k}\cdot\vec{J}$ and displays a band structure qualitatively similar to that of the $\Gamma$ point in RhSi (Fig.~2(d)).

\subsubsection{\large SM C.3 Fitted Parameters for RhSi}

\begin{figure}[t]
\includegraphics[width=140mm]{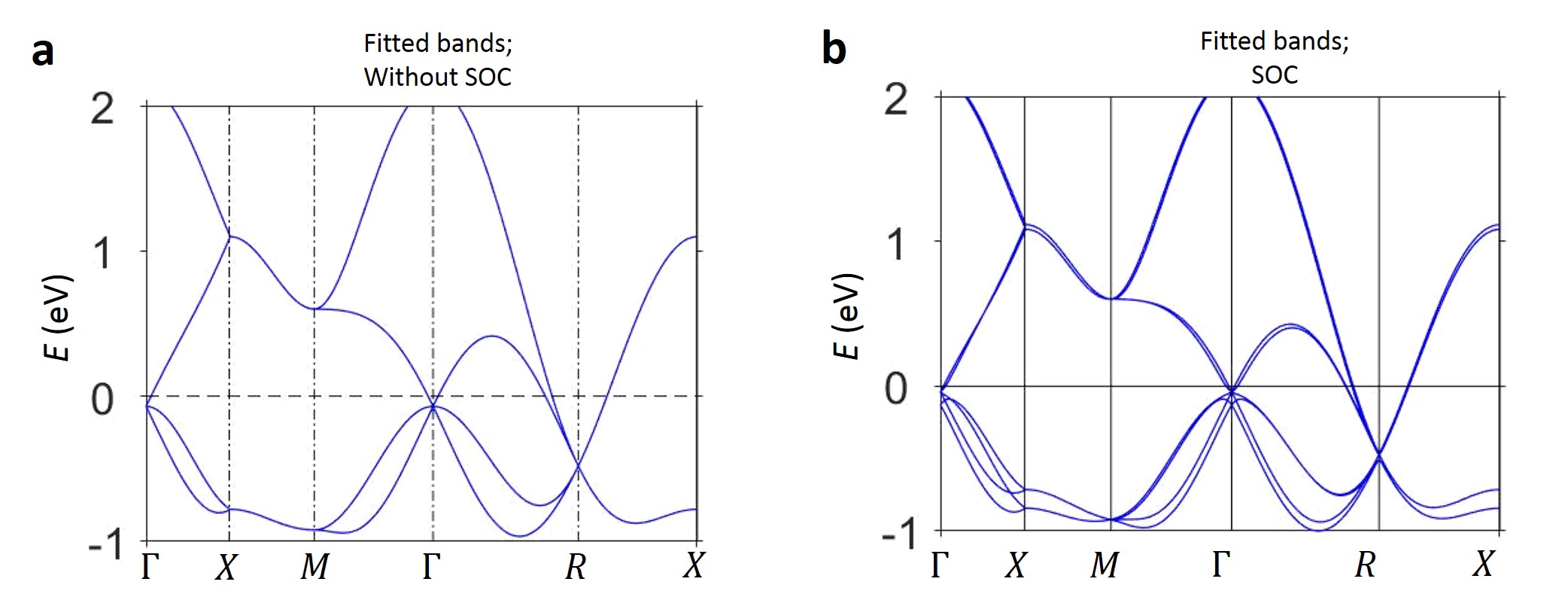}
\caption{\textbf{Bands obtained from tight-binding, fit to RhSi} (a) Bands without SOC. (b) Bands with SOC.}
\label{FigS5}
\end{figure}

Fitting the simple tight-binding model in Eq.~(\ref{eq:tb}) to the bands of RhSi near the Fermi energy, we find remarkably strong agreement given the simplicity of the model.  Using the eight parameters in Eq.~(\ref{eq:tb}), we are able to reproduce for RhSi all of the gaps and all of the unconventional fermion Chern numbers within those gaps.  In order to more adequately capture the energy offset between the fourfold fermion at $\Gamma$ and the double spin-1 Weyl at $R$, we add a ninth parameter, which represents an unsigned second-nearest-neighbor hopping between adjacent unit cells:
\begin{equation}
V_{2}(\vec{k})=v_{2}\bigg[\cos\left(k_{x}\right) + \cos\left(k_{y}\right) + \cos\left(k_{z}\right)\bigg]\mathds{1}
\end{equation}
where $\mathds{1}$ is the $8\times 8$ identity.  We obtain the parameters $v_{1}=0.55$, $v_{p}=-0.76$, $v_{2}=0.16$, $v_{r1}=0$, $v_{r2}=-0.03$, $v_{r3}=0.01$, $v_{s1}=-0.04$, $v_{s2}=0$ and $v_{s3}=0$ (Fig.~\ref{FigS5}), which in addition to reproducing all of the topological and band connectivity features of RhSi, also qualitatively describe bands closely resembling those calculated from first-principles DFT (Fig.~1(d)).

\subsection{\large SM D. Photocurrent Calculation Details}

To obtain a measure of the circular photogalvanic effect photocurrent rate, we calculate the multiband gyrotropic tensor $\beta_{ij}(\omega)$ as described in Ref.~\onlinecite{photogalvanic}:
\begin{equation}
\beta_{ij}(\omega)=\frac{i\pi e^{3}}{\hbar}\epsilon_{jkl}\sum_{\bold{k},n,m}f^{\bold{k}}_{nm}\Delta^{i}_{\bold{k},nm}r^{\bold{k}}_{k,nm} r^{l}_{\bold{k},mn}\delta(\hbar\omega - E_{\bold{k},mn})
\end{equation}
where $E_{\bold{k},nm}$ is the difference between the energies of bands $n$ and $m$ at crystal momentum $\bold{k}$, $f^{\bold{k}}_{nm}$ is the difference between the Fermi-Dirac distribution functions for bands $n$ and $m$ at $\bold{k}$, $r_{\bold{k},nm}=i\langle n|\partial_{\bold{k}}|m\rangle$ is the cross-gap Berry connection, and $\Delta^{i}_{\bold{k},nm}=\partial_{\bold{k}_{i}}E_{\bold{k},nm}/\hbar$.  We then trace over this tensor to find an effective photocurrent rate $dj/dt$; we note that the individual components of the photocurrent rate $dj_{i}/dt$ are still generically anisotropic for the fourfold-degenerate fermion near $\Gamma$ in RhSi.  To calculate the rates in Fig.~4(b), we tune to half-filling the tight-binding-derived $k\cdot p$ theories for a Kramers Weyl and for the fourfold unconventional fermion at $\Gamma$ in SG 198 ($t_{i}^{1}=t_{i}^{s}=0.04$ in Ref.~\onlinecite{KramersWeyl} and Eq.~(\ref{eq:gammakp}) with the fitted tight-binding parameters from SM C.3, respectively).  To produce the more realistic traced photocurrent rate in Fig.~4(d), we tune the Fermi energy to be slightly below half-filling in Eq.~(\ref{eq:gammakp}).  This rate also appears to saturate in the $k\cdot p$ regime at a value proportional to the Chern number $+4$ characterizing the finite-$q$ gap at the Fermi energy, though the individual components of $\beta_{ij}$ are still quite anisotropic.  By roughly calculating the interband matrix elements for circularly polarized light, we confirm that transitions are effectively disallowed between bands unless they are characterized by $\Delta m_{j}=\pm 1$, in accordance with the association of $m_{j}=\pm 1$ for right- or left-handed circularly polarized light~\cite{selectionrules}.

\clearpage

\subsection{\large SM E. Comparison of RhSi With Previous Examples of Weyl Semimetals}

In this section, we compare RhSi with previous proposed Weyl semimetal states from the following  perspectives: energy window of nontrivial topology, energy offset between chiral fermions, and Weyl point momentum-space separation.

\subsubsection{ \large E.1 Topologically Nontrivial Energy Window}

\begin{figure}[t]
\includegraphics[width=140mm]{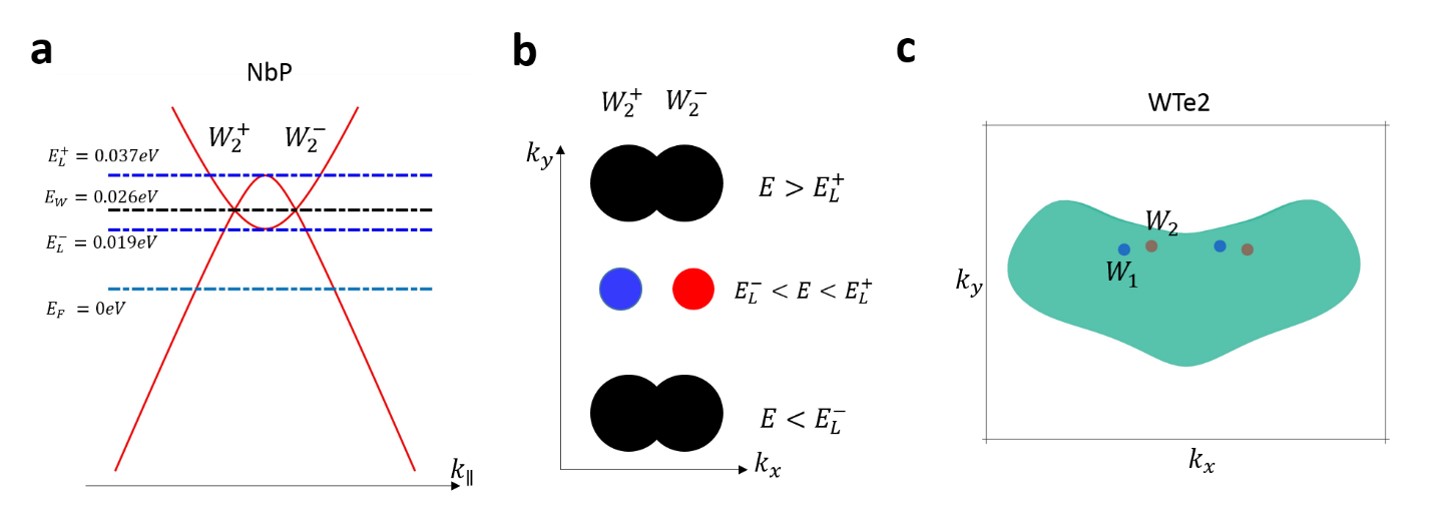}
\caption{\textbf{Fermi  surface topology of NbP and WTe$_{2}$} (a) The electronic structures of  the $W_{2}$ Weyl nodes in NbP. The  $W_{2}$ nodes sit around 26meV above Fermi level. The top and bottom Lifshitz transitions are around 37meV and 19meV above the Fermi level. (b)  A cartoon of the Fermi surface topology of NbP. The  red and blue colors indicate Fermi pockets of opposite chiral charge. The black pockets  contain no net chiral charge. (c) The Fermi pockets of  WTe$_2$  contain, respectively, chiral charges of net 0. Therefore, WTe$_{2}$ is  topologically trivial at the Fermi level.}
\label{FigS6}
\end{figure}

Weyl fermions and other topological chiral fermions are {\bf local} singularities in momentum space that can be regarded as sources or sinks of Berry curvature. The nontrivial topology of these chiral fermions  persists only in limited energy windows  where Fermi pockets  exhibit quantized non-zero chiral  charge. We emphasize that the isolation of Fermi pockets of opposite chiral charge is a key prerequisite for observing any topological phenomena associated with chiral fermions (such as the chiral anomaly and the quantized circular photogalvanic current). If the Fermi pockets arising from topological fermions with opposite chiral charges merge at the Fermi level, or if large trivial Fermi pockets with zero Chern number dominate the low-energy physics, then the novel physics of unconventional topological fermions cannot be isolated and studied. In previous works, the nontrivial energy window is closely related to the degree of band inversion \cite{arcDetect1}, which in previous works has been quite narrow (around 10-100meV. See Table S1). For example, the  topologically nontrivial energy window in NbP is around 20meV~\cite{Huang2015, Weng2015}. The energy dispersion passing through $W_{2}$ in NbP is plotted in Fig.~\ref{FigS6}(a), and the cartoon of Fermi pockets at different energies are plotted in Fig.~\ref{FigS6}(b). Within the Lifshitz transitions, the two compensative $W_{2}$ are enclosed in two isolated pockets. In contrast,  away from the Lifshitz transitions, the two pockets merge into one trivial pocket. Since the Fermi level of NbP is  away from the Lifshitz transitions,  NbP is topologically trivial at the Fermi energy. Similarly, as shown in Fig.~\ref{FigS6}(c), WTe$_{2}$ is also topological trivial at the Fermi level.

\begin{figure}[t]
\includegraphics[width=140mm]{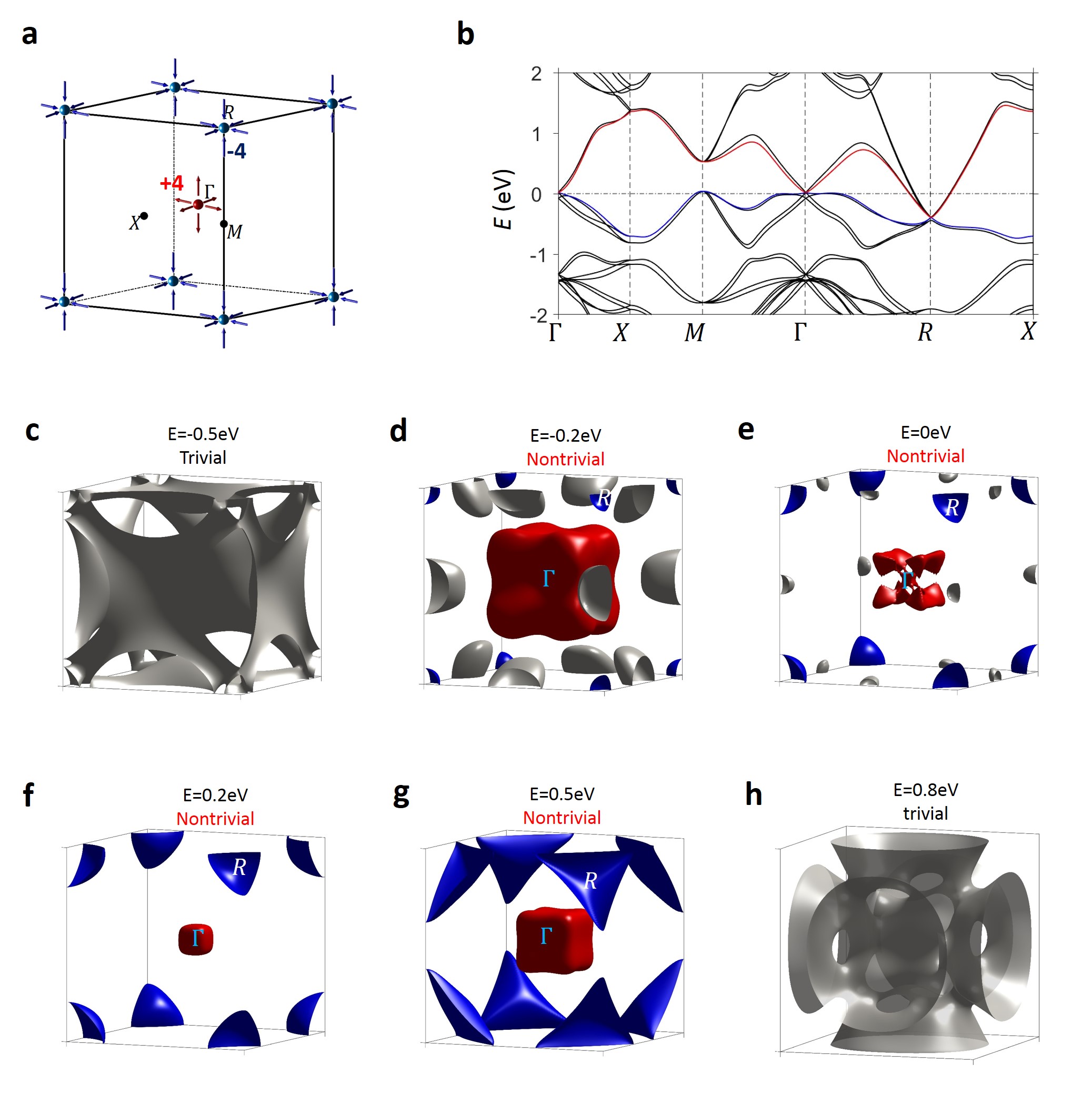}
\caption{\textbf{Fermi surface topology of RhSi} (a) The distribution of chiral  charge in RhSi between the conduction and valence bands. (b) The electronic structure of RhSi. (c-h) Topology of RhSi Fermi surfaces at different energies. The red pockets contain  chiral charge $+4$, the blue ones contain charge $-4$, and the black ones are topologically trivial. In (d-g), the red and blue pockets are  isolated, and therefore RhSi is topologically nontrivial at these energies. In (e) and (h), the Fermi pockets enclosing  $\Gamma$ and $R$ mix with each other, forming a  topologically trivial pocket.}
\label{FigS7}
\end{figure}

In contrast, RhSi has a much larger topologically nontrivial energy window  of around 1.2eV. Figure~\ref{FigS7}(a) illustrates the  chiral charge distribution of RhSi in the Brillouin zone. The Fermi pockets in RhSi at  a range of energies are plotted in  Figs.~\ref{FigS7}(c-h), in which the red pockets enclosing $\Gamma$ contain charge $+4$, the blue pockets enclosing $R$ contain charge $-4$, and the black pockets are topologically trivial.  The red and blue pockets remain  clearly isolated  at energies of -0.2eV, 0eV, 0.2eV and 0.5eV (Figs.~\ref{FigS7}(d-g)), and merge into one trivial pocket at -0.5eV and 0.8eV (Figs.~\ref{FigS7}(c,h)). This energy window is extremely large, and therefore suggests that the topological properties of RhSi should be quite robust to impurities, implying that RhSi is an ideal platform for topological and chiral experimental applications, even at room temperature.

 \begin{table}
\begin{center}
\centering
\begin{scriptsize}
\begin{tabular}{p{2cm}p{3cm}p{1.8cm}p{1.8cm}p{4.5cm}p{1.8cm}}
\hline
 Materials & $k$-separation ($\textrm{\AA}^{-1}$)   & Number & Charge & Topological Energy Window & $\Delta E$ (eV)\\
\hline
RhSi & $\sim$2.33 & 2 & 4 & $\sim$1.2eV; nontrivial at EF=0 & $\sim$0.41 \\
TaAs \cite{Huang2015, Weng2015} & $\sim$0.07 & 24 & 1 &  $\sim$0.08eV; nontrivial at EF=0 & 0 \\
TaP \cite{Huang2015, Weng2015} & $\sim$0.06  & 24 & 1 & $\sim$0.07eV; nontrivial at EF=0 & 0 \\
NbAs \cite{Huang2015, Weng2015} & $\sim$0.02 & 24 & 1 & $\sim$0.02eV; nontrivial at EF=0 & 0\\
NbP \cite{Huang2015, Weng2015} & $\sim$0.02 & 24 & 1 & $\sim$0.02eV; Trivial at EF=0 & 0\\
WTe$_2$ \cite{type2Weyl1}& $<$0.01 & 8 & 1 & $\sim$0.01eV; Trivial at EF=0  & 0\\
LaAlGe \cite{LaAlGe} & $<$0.01 & 40 & 1 & nontrivial at EF=0 & 0\\
\hline
\end{tabular}
\end{scriptsize}
\end{center}

\caption{Comparison between RhSi and previous works}
\end{table}

\clearpage
\subsubsection{ \large E.2  Energy Difference between Compensative Charges}
\begin{figure}[t]
\includegraphics[width=140mm]{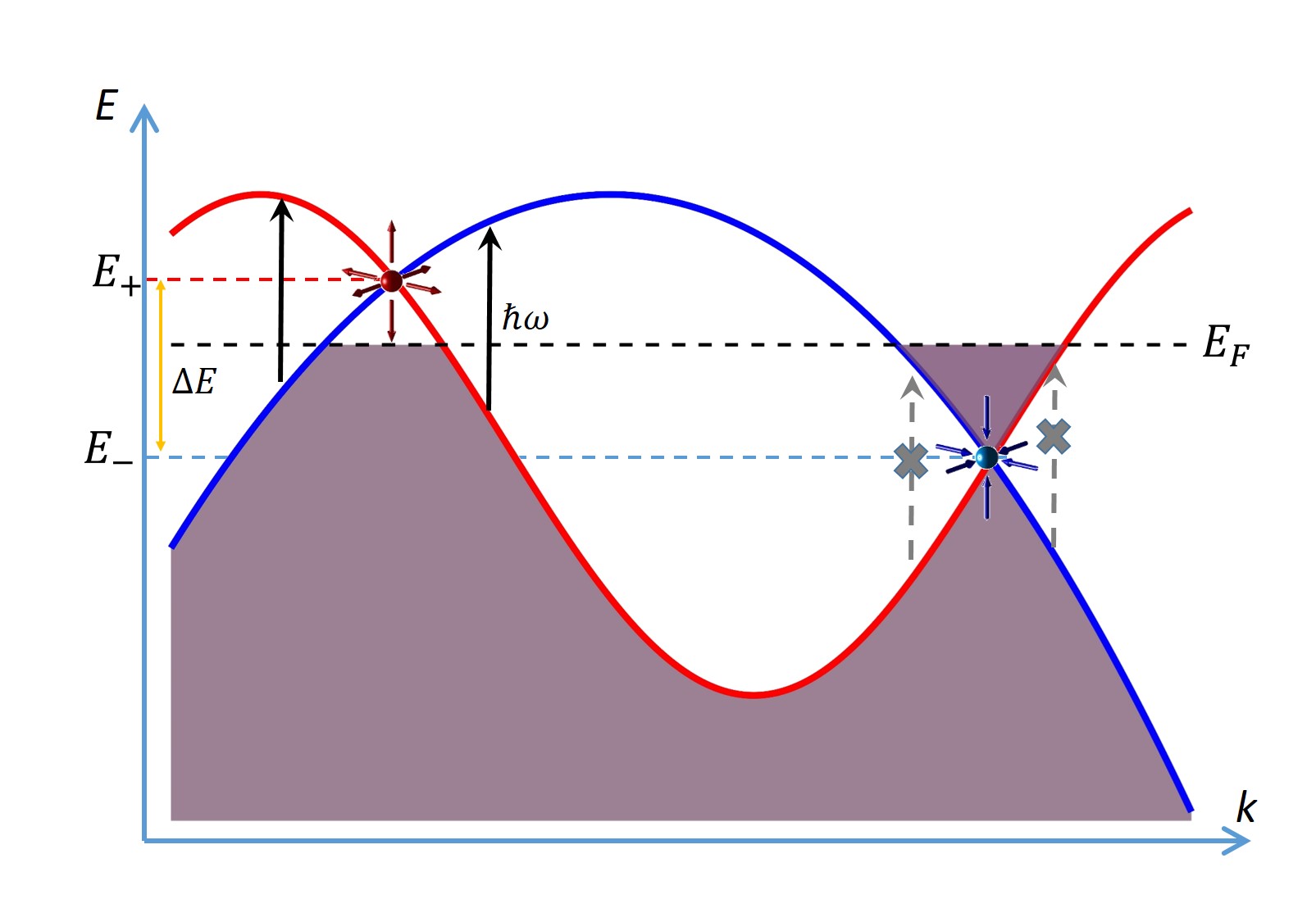}
\caption{\textbf{Quantized quantized circular photogalvanic effects require the energy offset between positive and negative charges.}}
\label{FigS8}
\end{figure}

 All of the previous examples of Weyl semimetals relied on band inversion in crystals with mirror symmetries.  Consequently, their Weyl points always had partners related to each other by the mirrors, and they were forced to sit pairwise at the same energies.  In RhSi, the chiral fermions sit at the time-reversal-invariant momenta, and therefore are not related to each other by \emph{any} symmetries.  Consequently, they are free to sit at different energies.  The cartoon in Figure~\ref{FigS8} shows the energy difference between compensative chiral charges  in several Weyl semimetals.

 In particular, this energy offset has been identified as a necessary ingredient for a range of topological and chiral transport effects. For example, an external magnetic field  has been proposed to induce an equilibrium dissipationless current when opposite chiral fermions are at different energies \cite{Weylcurrent},  with the amplitude of the induced  current being proportional to the energy difference between compensative charges.  The 0.41eV energy  offset between oppositely charged chiral fermions in RhSi  renders it an ideal platform  for realizing this  effect.   A chiral-node energy offset also allows the realization of the quantized photogalvanic effect \cite{photogalvanic}, discussed here in both the conclusion of the main text and in further detail in SM D.

\subsubsection{ \large E.3 Momentum-Space Weyl Node Separation and Fermi Arc Length}
The separation of compensative Weyl fermions in momentum space is one criterion to evaluate the robustness of Weyl semimetals.  In previous works,  this separation has been quite small, on the order of $10^{-2}$ $\textrm{\AA}^{-1}$.  In contrast, in RhSi, the compensative chiral fermions  have a separation of 2.33 $\textrm{\AA}^{-1}$. The detailed data are shown in Table S1.

\begin{figure}[t]
\includegraphics[width=140mm]{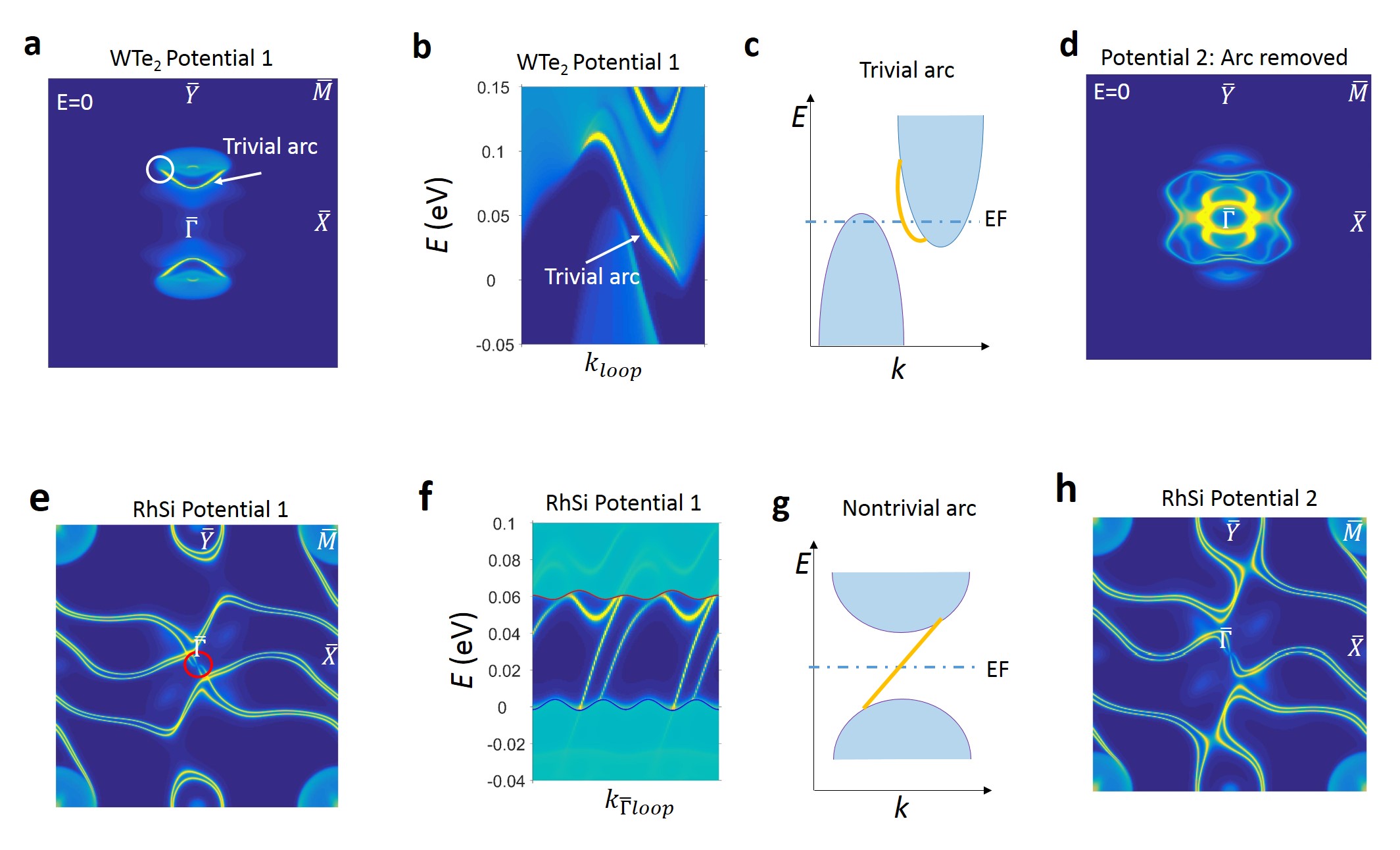}
\caption{\textbf{Trivial arcs in WTe$_{2}$ and nontrivial Fermi arcs in RhSi} (a) The trivial Fermi arc of WTe$_{2}$. (b) The energy dispersion of  the surface states of WTe$_{2}$ along the  white loop. The arc connecting  the conduction band in WTe$_{2}$ to itself is trivial. (c)  A cartoon of the  dispersion of the trivial arc in WTe$_{2}$. The trivial arc (yellow line) could be moved into the  projection of the bulk band manifold.  (d) The surface states of WTe$_{2}$ after increasing the surface chemical potential. The trivial arc is removed under the new surface potential. (e) The  topologically nontrivial Fermi arcs in RhSi cross over the whole surface BZ. (f) The energy dispersion of the topological Fermi arcs along the red path. (g)  A cartoon of topological Fermi arcs;  nontrivial Fermi arcs connecting the conduction and valence bands are protected by the  bulk Chern numbers, and cannot be removed without merging bulk nodes. (h) The surface states of RhSi after increasing the surface chemical potential. The topological Fermi arcs are still preserved.}
\label{FigS9}
\end{figure}

 The surface consequences of these bulk Weyl nodes, topological surface Fermi arcs, span a distance constrained by the separation of the bulk nodes.  The longest Fermi arcs thus far experimentally observed have been in TaAs~\cite{Huang2015, Weng2015}, and are still significantly shorter than the zone-spanning ones we propose in RhSi.  Large Fermi arcs crossing a quarter of the surface BZ have also been observed in WTe$_{2}$; however they have been shown to be topologically trivial~\cite{trivialarc, LaAlGe}, and thus highly sensitive to surface conditions.  In particular, by changing the surface chemical potential, the arcs in WTe$_{2}$ could be fully removed (Fig.~\ref{FigS9}(a-d)).  In contrast, the long Fermi arcs in RhSi are topologically protected by the bulk chiral charge of unconventional Weyl  fermions, and will therefore will be robust to surface conditions (Fig.~\ref{FigS9}(e-h)).

To summarize, RhSi is by far the most ideal candidate topological chiral fermion semimetal thus far identified.  In Table S1, we provide a comparison of RhSi with previously proposed Weyl semimetals, the results of which can be summarized as:

      (1)	RhSi has at its Fermi energy the {\bf minimum} number of isolated chiral Fermions (2) allowed in a BZ.

      (2)	The two topological fermions have the {\bf largest} possible separation in momentum space.

      (3)	RhSi has the {\bf longest} possible topological Fermi arcs.

      (4)	RhSi has the {\bf largest} chiral charge ever featured in a realistic material (4 times that of previous works).

      (5)	RhSi has the {\bf largest}  topologically nontrivial energy window  of any material thus far.

\clearpage

\subsection{\large SM F.  Numerical Calculations of the Chiral Charges of High-fold Fermions}
\begin{figure}[t]
\includegraphics[width=140mm]{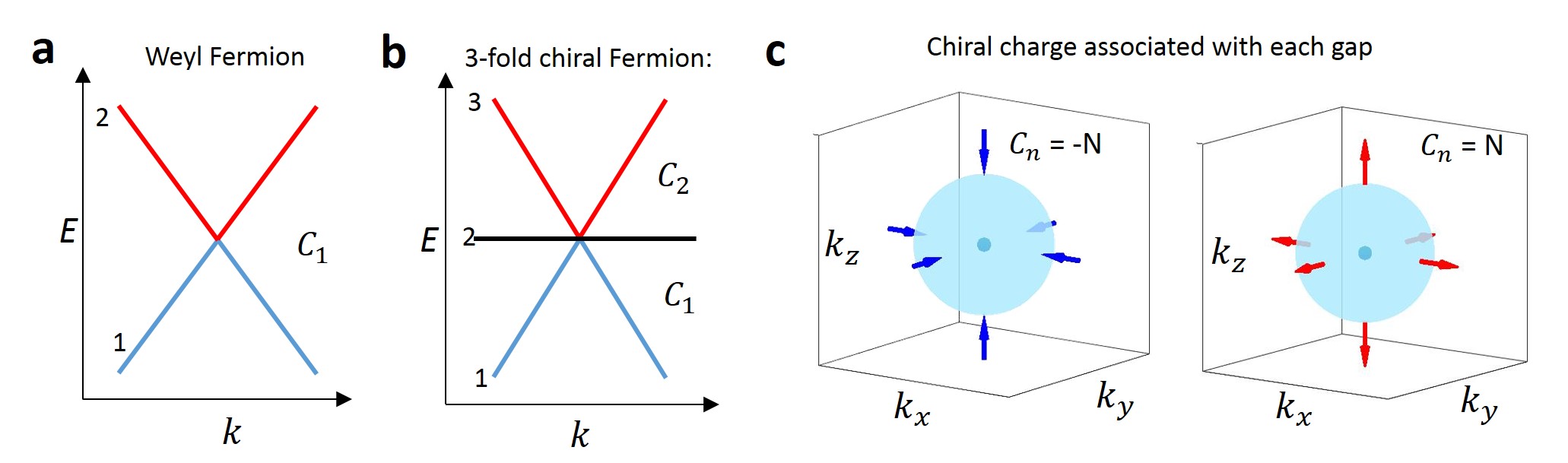}
\caption{\textbf{Chirality of high-fold fermions} (a) A Weyl fermion with one band gap has a quantized Chern number. (b) A 3-fold fermion has two Chern numbers ($C_{1}$,$C_{2}$) associated with the two band gaps. (c) In general, an N-fold fermion with N-1 band gaps has well defined N-1 Chern numbers ($C_{1}$,$C_{2}$,...,$C_{N-1}$) via the integration up to different bands.}
\label{FigS10}
\end{figure}

A conventional Weyl fermion  forms at the degeneracy of two singly degenerate bands with one band gap. The chirality of a Weyl fermion is calculated from  the Berry curvatures of  the occupied bands,  and dictates the  minimum number of nontrivial surface states in the gap between the conduction and valence bands around the Weyl node.  A cartoon of a Weyl fermion is  drawn in Fig.~\ref{FigS10}(a). The chiral charge associated with the gap is labeled as $C_{1}$.  In general, an N-fold fermion is formed by N bands with N-1 band gaps. For example, a  threefold fermion is formed by three bands with two band gaps around, as shown in Fig.~\ref{FigS10}(b).  For a Weyl fermion with multiple band gaps, we can calculate the chiral charge around each gap.  For example, if we set the top valence band as the blue band (band 1),  we observe a Chern number $C_{1}$, which decides the number of Fermi arcs in the gap between band 1 (blue) and band 2 (red). Similarly, if we set band 1 and 2 as the valence bands,  we observe a different Chern number associated with the gap between band 2 (black) and band 3 (red), which is labeled as $C_{2}$. Therefore, the chiral charges of a  threefold fermion  can be defined by two numbers ($C_{1}$,$C_{2}$). In general, for any  N-fold isolated fermion, we  can calculate N-1  chiral charges ($C_{1}$,$C_{2}$,...,$C_{N-1}$) associated with different  band occupations. These values decide the number of Fermi arcs in each band gap around the high-fold chiral fermion.

\subsection{\large SM G. The Effects of Exchange-correlation Pseudopotentials on RhSi}
\begin{figure}[t]
\includegraphics[width=140mm]{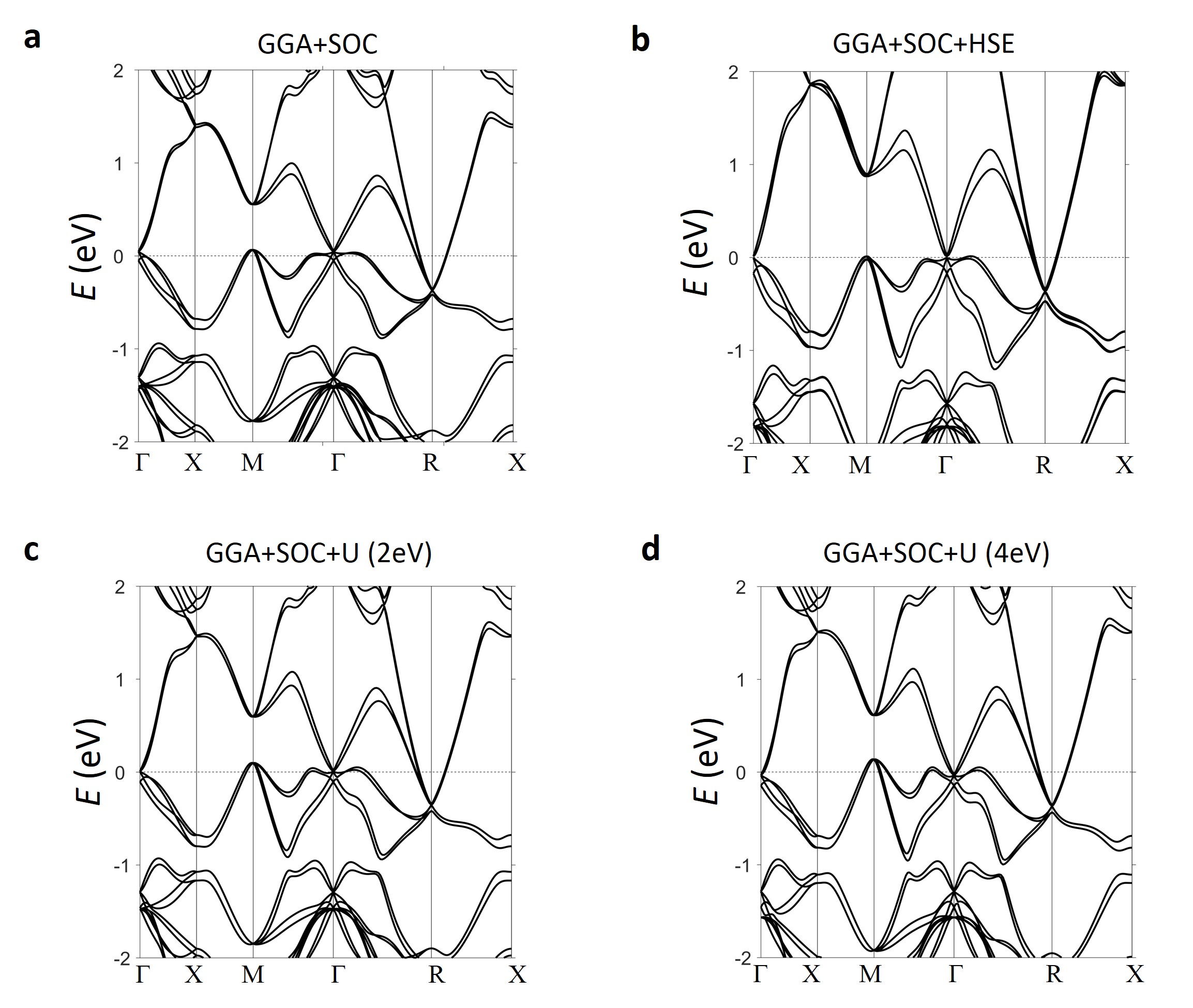}
\caption{\textbf{Electronic structures of RhSi with different  exchange-correlation potentials} (a) Electronic structures of RhSi. (b) Electronic structures of RhSi after band gap corrections. (c,d) Band structures of RhSi when U=2eV and 4eV, respectively.}
\label{FigS11}
\end{figure}

Here we take the effects of exchange-correlation pseudopotentials into consideration. Specifically, we have conducted generalized gradient approximation (GGA) calculations with the Hubbard energies U=2eV and U=4eV. We have also done the calculation with the Heyd–Scuseria–Ernzerhof (HSE) exchange-correlation functional. The electronic structures of RhSi under different exchange-correlation pseudopotentials are shown in Fig.~\ref{FigS11}. The fourfold- and sixfold-degenerate fermions are still preserved and their Fermi pockets are still remained well separated in momentum space. This is because the fourfold and sixfold fermions in our work are symmetry guaranteed. And the exchange-correlation potentials do not break the crystal symmetries.

\begin{thebibliography}{99}
\bibitem{Weyl} H. Weyl, I. Z. Phys. $\mathbf{56}$, 330 (1929).
\bibitem{Wilczek} F. Wilczek, Phys. Today $\mathbf{51}$, 11 (1998).
\bibitem{Volovik2003} G. E. Volovik,  The Universe in a Helium Droplet (Clarendon Press, Oxford, 2003).

\bibitem{NewFermion} B.  Bradlyn \textit{et al.}, Science  $\mathbf{353}$, aaf5037 (2016).

\bibitem{Kane05p226801} C.  L.  Kane  and  E.  J.  Mele,  Phys.  Rev.  Lett.  $\mathbf{95}$,  226801  (2005).
\bibitem{Kane05p146802} C.  L.  Kane  and  E.  J.  Mele,  Phys.  Rev.  Lett.  $\mathbf{95}$,  146802  (2005).
\bibitem{Hasan2010} M.  Z.  Hasan  and  C.  L.  Kane,  Rev.  Mod.  Phys.  $\mathbf{82}$,  3045  (2010).
\bibitem{Qi2011} X.-L.  Qi  and  S.-C.  Zhang,  Rev.  Mod.  Phys.  $\mathbf{83}$,  1057  (2011).
\bibitem{BernevigHgTe} B.  A.  Bernevig,  T.  L.  Hughes,  and  S.-C.  Zhang,  Science  $\mathbf{314}$,  1757  (2006).
\bibitem{RevBansil} A.  Bansil,  H.  Lin,  and  T.  Das,  Rev.  Mod.  Phys.  $\mathbf{88}$,  021004  (2016).

\bibitem{Wan2011} X.  Wan \textit{et al.},  Phys.  Rev.  B  $\mathbf{83}$,  205101  (2011).
\bibitem{Burkov2011} A.  A.  Burkov  and  L.  Balents,  Phys.  Rev.  Lett.  $\mathbf{107}$,  127205  (2011).
\bibitem{Murakami2007}  S.  Murakami,  New  Journal  of  Physics  $\mathbf{9}$,  356  (2007).
\bibitem{Huang2015}  S.-M.  Huang,  S.-Y.  Xu \textit{et al.},  Nature  Communications  $\mathbf{6}$,  7373 (2015).
\bibitem{Weng2015} H.  Weng \textit{et al.},   Phys.  Rev.  X   $\mathbf{5}$,  011029  (2015).
\bibitem{Hasan_TaAs} S.-Y.  Xu  \textit{et al.},  Science  $\mathbf{349}$,  613  (2015).
\bibitem{TaAs_Ding} B.  Q.  Lv \textit{et al.},  Phys.  Rev.  X  $\mathbf{5}$,  031013  (2015).
\bibitem{MIT_Weyl} L.  Lu \textit{et al.}, Science  $\mathbf{349}$,  622 (2015).

\bibitem{type2Weyl1} A.  A.  Soluyanov \textit{et al.},  Nature $\mathbf{527}$,  495 (2015).
\bibitem{type2Weyl2} Y. Xu, F. Zhang and C. Zhang,  Phys.  Rev.  Lett.  $\mathbf{115}$, 265305  (2015).
\bibitem{LaAlGe} S.-Y.  Xu  \textit{et al.}, Science Advances, $\mathbf{3(6)}$, e1603266 (2017).
\bibitem{KramersWeyl} G.  Chang \textit{et al.}, Preprint at https://arxiv.org/abs/1611.07925 (2016).
\bibitem{DoubleDirac} B.  J.  Wieder,  Y.  Kim,  A.  M.  Rappe,  and  C.  L.  Kane,  Phys.  Rev.  Lett.  $\mathbf{116}$,  186402  (2016).



\bibitem{trip1} H.  Weng,  C.  Fang,  Z.  Fang,  and  X.  Dai,  Phys.  Rev.  B  $\mathbf{93}$,  241202  (2016).
\bibitem{trip2} Z.  Zhu,  G.  W.  Winkler,  Q.  Wu,  J.  Li,  and  A.  A.  Soluyanov,  Phys.  Rev.  X  $\mathbf{6}$,  031003  (2016).
\bibitem{Nexus}  G.  Chang \textit{et al.}, Scientific Reports $\mathbf{7}$, 1688 (2017).
\bibitem{Phonon}  T. Zhang, Z. Song, A. Alexandradinata, H. Weng, C. Fang, L. Lu, Z. Fang, Preprint at https://arxiv.org/abs/1705.07244 (2017).

\bibitem{MMonFS} S.  Zhong,  J.  E.  Moore,  and  I.  Souza,  Phys.  Rev.  Lett.  $\mathbf{116}$,  077201  (2016).
\bibitem{chiral_mag} J.  Ma  and  D.  A.  Pesin,  Phys.  Rev.  B  $\mathbf{92}$,  235205  (2015).
\bibitem{photocurrentweyl} C.-K. Chan \textit{et al.}, Phys. Rev. B $\mathbf{95}$, 041104 (R) (2017).
\bibitem{photoTaAs} Q. Ma, S.-Y.  Xu  \textit{et al.}, Nat. Phys. $\mathbf{13}$, 842 (2017).
\bibitem{photogalvanic} F.  de  Juan \textit{et al.}, Nat. Comm. $\mathbf{8}$, 15995 (2017). 
\bibitem{arc1} L.  Balents,  Physics  $\mathbf{4}$,  36  (2011).
\bibitem{arc2} T.  Ojanen,  Phys.  Rev.  B  $\mathbf{87}$,  245112  (2013).
\bibitem{Riemann} C.  Fang,  L.  Lu,  J.  Liu,  and  L.  Fu,  Nat.  Phys.  $\mathbf{12}$,  936  (2016).

\bibitem{Nexus_exp} B. Q. Lv \textit{et al.}, Nature $\mathbf{546}$, 627 (2017).
\bibitem{RhSi} I.  Engstr$\ddot{o}$m  and  J.  Torsten,  Acta  Chemica  Scandinavica  $\mathbf{19}$,  1508  (1965).


\bibitem{fillingconstraint}  H.  Watanabe,  H.  C.  Po,  A.  Vishwanath,  and  M.  Zaletel,  PNAS  $\mathbf{112}$,  14551  (2015).
\bibitem{QuantumChemistry}  B.  Bradlyn \textit{et al.},  Nature $\mathbf{547}$,  298  (2017).

\bibitem{magDirac2} S.  M.  Young  and  B.  J.  Wieder,  Phys.  Rev.  Lett.  $\mathbf{118}$,  186401  (2017).



\bibitem{BigBook} C.  J.  Bradley  and  A.  P.  Cracknell,  The Mathematical Theory of Symmetry in Solids  (Clarendon
Press  Oxford,  Oxford,  United  Kingdom,  1972),  ISBN  0199582580.

\bibitem{Layer_group} B.  J.  Wieder  and  C.  L.  Kane,  Phys.  Rev.  B  $\mathbf{94}$,  155108  (2016).

\bibitem{Adrian19} A.  Bouhon  and  A.  Black-Schaffer, Preprint at https://arxiv.org/abs/ 1702.05343 (2017).
\bibitem{BalatskySG19}R.  M.  Geilhufe,  S.  S.  Borysov,  A.  Bouhon,  and  A.  V.  Balatsky, Preprint at https://arxiv.org/abs/ 1611.04316 (2017).

\bibitem{trivialarc} F.Y. Bruno \textit{et al.},  Phys.  Rev.  B $\mathbf{94}$, 121112(R) (2016).

\bibitem{arcDetect2} S.-Y.  Xu  \textit{et al.},  Phys.  Rev.  Lett.   $\mathbf{116}$,  096801  (2016).

\bibitem{spintronics1} I.  $\check{Z}$uti$\acute{c}$,  J.  Fabian,  and  S.  Das  Sarma,  Rev.  Mod.  Phys.  $\mathbf{76}$,  323  (2004).
\bibitem{spintronics2} A.  R.  Mellnik \textit{et al.},  Nature  $\mathbf{511}$,  449  (2014).
\bibitem{selectionrules}  C.  L.  Tang  and  H.  Rabin,  Phys.  Rev.  B  $\mathbf{3}$,  4025  (1971).


\bibitem{CoSi_zhang}  P. Tang, Q. Zhou, S-C. Zhang, Preprint at https://arxiv.org/abs/1706.03817 (2017).









\bibitem{openmx1} T.  Ozaki,  Phys.  Rev.  B  $\mathbf{67}$,  155108  (2003).
\bibitem{openmx2} T.  Ozaki  and  H.  Kino,  Phys.  Rev.  B  $\mathbf{69}$,  195113  (2004).
\bibitem{DFT2} G.  Kresse  and  J.  Furthm¨uller,  Phys.  Rev.  B  $\mathbf{54}$,  11169  (1996).
\bibitem{DFT3} G.  Kresse  and  D.  Joubert,  Phys.  Rev.  B  $\mathbf{59}$,  1758  (1999).
\bibitem{DFT4} J.  P.  Perdew,  K.  Burke,  and  M.  Ernzerhof,  Phys.  Rev.  Lett.  77,  3865  (1996).

\bibitem{CoSi}  P.  Demchenko \textit{et al.}, Chem.  Met.  Alloys  $\mathbf{1}$,  50  (2008).
\bibitem{BaPtP} G.  Wenski  and  A.  Mewis,  Zeitschrift  f$\ddot{u}$r  anorganische  und  allgemeine  Chemie  $\mathbf{535}$,  110  (1986).

\bibitem{CoGe}  H.  Takizawa,  T.  Sato,  T.  Endo,  and  M.  Shimada,  Journal  of  Solid  State  Chemistry  $\mathbf{73}$, 40  (1988).
\bibitem{RhGe}  V.  Larchev  and  S.  Popova,  Journal  of  the  Less  Common  Metals  $\mathbf{87}$,  53  (1982).
\bibitem{AlPd}  M.  Ettenberg,  K.  L.  Komarek,  and  E.  Miller,  Metallurgical  Transactions  $\mathbf{2}$,  1173  (1971).
\bibitem{AlPt} K.  Schubert \textit{et al.},  Naturwissenschaften  $\mathbf{43}$,  248  (1956).

\bibitem{FeSi} S.  Ono,  T.  Kikegawa,  and  Y.  Ohishi,  European  journal  of  mineralogy  $\mathbf{19}$,  183  (2007).
\bibitem{GaPd} E.  Hellner  and  F.  Laves,  Zeitschrift  Naturforschung  Teil  A  $\mathbf{2}$,  177  (1947).

\bibitem{GaPt} A.  Marc,  B.  Horst,  W.  Michael,  P.  Yurii,  G.  Rainer,  and  G.  Peter,  $\mathbf{225}$,  617  (2010).
\bibitem{WilsonLoop} L.  Fidkowski,  T.  Jackson,  and  I.  Klich,  Phys.  Rev.  Lett.  $\mathbf{107}$,  036601  (2011).
\bibitem{arcDetect1} I.  Belopolski \textit{et al.}, Phys.  Rev.  Lett.  $\mathbf{116}$,  066802  (2016).

\bibitem{Weylcurrent} A. A. Zyuzu, S. Wu and A. A. Burkov, Phys.  Rev.  B  $\mathbf{85}$,  165110  (2012).
\bibitem{trivialarc} F.Y. Bruno \textit{et al.},  Phys.  Rev.  B $\mathbf{94}$, 121112(R) (2016).











%
















\end{thebibliography}
\end{document}